# Observing the evolution of the Sun's global coronal magnetic field over eight months

Zihao Yang[1,2], Hui Tian[1,3]*, Steven Tomczyk[2,4], Xianyu Liu[1,5], Sarah Gibson[2], Richard J. Morton[6], Cooper Downs[7]

[1]School of Earth and Space Sciences, Peking University; Beijing, 100871, People's Republic of China.

[2]High Altitude Observatory, National Center for Atmospheric Research; Boulder, CO, 80307, USA.

[3]Key Laboratory of Solar Activity and Space Weather, National Space Science Center, Chinese Academy of Sciences; Beijing 100190, People's Republic of China.

[4]Solar Scientific LLC; Boulder, CO, 80301, USA.

[5]Department of Climate and Space Sciences and Engineering, University of Michigan; Ann Arbor, MI, 48109, USA.

[6]Department of Mathematics, Physics and Electrical Engineering, Northumbria University; Newcastle Upon Tyne NE1 8ST, UK.

[7]Predictive Science Inc.; San Diego, CA 92121, USA.

*Corresponding author. Email: huitian@pku.edu.cn

**The magnetic field in the Sun's corona stores energy that can be released to heat the coronal plasma and drive solar eruptions. Measurements of the global coronal magnetic field have been limited to a few snapshots. We present observations using the Upgraded Coronal Multi-channel Polarimeter, which provided 114 magnetograms of the global corona above the solar limb spanning approximately eight months. We determined the magnetic field distributions at different solar radii in the corona, and monitored the evolution at different latitudes over multiple solar rotations. We found varying field strengths from <1 to 20 Gauss within 1.05-1.6 solar radii and a signature of active longitude in the coronal magnetic field. Coronal models are generally consistent with the observational data, with larger discrepancies in high-latitude regions.**

The magnetic field plays a dominant role in many physical processes occurring on the Sun, including transient solar eruptions, plasma heating and the 11-year solar activity cycle. Understanding these phenomena requires measurements of the evolving magnetic field at all altitudes in the Sun's atmosphere (*1-3*). Magnetic fields on the Sun's visible surface (the photosphere) are routinely measured using the Zeeman effect, the splitting of some spectral lines due to the presence of an external magnetic field. However, it is difficult to extend such measurements to the Sun's upper atmosphere, particularly the corona, due to the much weaker emission and smaller line splitting. The coronal magnetic field is the source of energy that heats the hot plasma in the corona (*4*) and drives intermittent solar eruptions (*5*), but no routine measurements of the magnetic field in the corona are available.

Some individual measurements of the coronal magnetic field have been made. For example, spectro-polarimetric measurements of infrared spectral lines have been used to determine the magnetic fields in specific regions of the corona with very strong field (*6-8*). Radio spectral imaging observations have been used to estimate coronal magnetic field strengths in flaring regions (*9-11*). A predicted relationship between the intensity of the Fe X 25.7 nm ultraviolet spectral line intensity and the magnetic field strength (*12*), has been proposed as a means to measure the coronal magnetic field (*13*); however the precision required by that technique is not available with existing instruments (*14*). An alternative technique is coronal seismology, which has been used to infer magnetic field strengths from observations of magnetohydrodynamics (MHD) waves in the corona (*15-17*). However, this technique is usually applied to single oscillation events, which provide only a single value for the field strength within an oscillating structure. None of these techniques have been routinely implemented to monitor the evolution of the Sun's global coronal magnetic field.

Coronal seismology has been further applied to propagating transverse MHD waves (*18, 19*), which are more pervasive in the corona than individual transient oscillations. Observations of the corona above the solar limb using the Coronal Multi-channel Polarimeter (CoMP) (*20*) have been used to produce individual global coronal magnetograms (maps of magnetic field strength and direction) (*21, 22*). This two-dimensional (2D) coronal seismology technique maps the plane-of-sky (POS) component of the global coronal magnetic field. However, due to the limited signal-to-noise ratio (S/N), only a few CoMP datasets have been used to generate global magnetograms. Routine monitoring would require approximately daily measurements of the coronal magnetic field.

**Observations of the corona**

We analyzed observations of the corona with the Upgraded Coronal Multi-channel Polarimeter (UCoMP) (*23*). Daily scientific observations with UCoMP began in mid-2021. Following completion of instrument commissioning, the data quality stabilized in early 2022. Observations were performed almost daily from that point until halted by a volcanic eruption of Mauna Loa (UCoMP is located on the slopes of that mountain) in late November 2022. UCoMP provides imaging spectral observations of the off-limb (beyond the solar disk) corona at all latitudes. Compared to CoMP, it has an expanded field-of-view (FOV) covering ~1.05 to 1.6 solar radii, a higher spatial resolution of ~6″, higher sensitivity and improved data quality stability.

UCoMP performs imaging spectroscopy by recording a spectrum at each pixel within the FOV, targeting the Fe XIII 1074.7 nm and 1079.8 nm near-infrared spectral lines. Each spectral profile was fitted with a Gaussian function to determine the line intensity and Doppler velocity of the coronal plasma at each location within the FOV (*24*). The intensity ratio of the two lines was used to map the electron number density (and therefore plasma mass density, assuming overall neutral charge) in the corona (*21, 22*).

Previous observations with CoMP have shown prevalent propagating disturbances in the Doppler velocity maps of Fe XIII 1074.7 nm, which were interpreted as being due to propagating transverse MHD waves, termed kink waves (*18, 19, 21*). Similar evidence for kink waves is present in the UCoMP data (movie S1). We modified (*25*) a previous wave tracking technique (*26*) and applied it to the UCoMP observations to determine the wave propagation direction and phase speed at each pixel within the FOV.

The data quality of the UCoMP observations was sufficient for us to derive maps of plasma density, wave propagation direction and phase speed in the global corona for 114 days during the 253-day (approximately 8-month) period of 19 February to 29 October 2022. Most data gaps are related to bad weather conditions. This period spanned more than nine rotations of the Sun (Carrington rotations, with a period of approximately 27.27 days). In contrast, previous CoMP observations provided usable maps on only one or two days in one year (*21, 22*).

**Determining the coronal magnetic field**

Using the density diagnostics and wave tracking results, we applied 2D coronal seismology to determine global maps of the coronal magnetic field (*21, 22*). Under the condition of magnetic pressure dominance (which applies in the corona) and at UCoMP's resolution of ~6″, the magnetic field strength is related to the local plasma density and phase speed of the observed kink wave as (*17, 25, 27*):

$$B = v_{\text{ph}}\sqrt{\mu_0 \rho} \qquad (1)$$

where $B$ is the magnetic field strength, $\rho$ is the plasma density, $v_{\text{ph}}$ is the wave phase speed, and $\mu_0$ is the magnetic permeability in a vacuum. Because the corona is optically thin (minimal attenuation) to the Fe XIII lines at 1074.7 nm and 1079.8 nm, the observed spectral profiles are integrations of the line emission along the line-of-sight (LOS). Therefore, the magnetic field strength derived using Equation 1 is a weighted average along the LOS.

We applied Equation 1 to the UCoMP global maps of plasma density and wave phase speed, to produce 114 global maps of coronal magnetic field (movie S2). Fig. 1 shows three examples of our results on 21 February, 1 June and 5 August 2022, along with simultaneous coronal images at 19.5 nm from the Solar UltraViolet Imager (SUVI) (*28*) on the Geostationary Operational Environmental Satellite. The magnetic field strengths in the UCoMP FOV are mostly in the range of 0.5 to 4 Gauss. In some cases, at low altitudes above active regions (regions with strong 19.5 nm emission), the coronal field we inferred reaches ~20 Gauss. We estimated the uncertainties on these field strengths to be generally <20% (*25*). Because the measured wave phase speeds are only the projection of the true speeds onto the POS, perpendicular to the LOS, the measured field strengths are the POS component of the coronal magnetic field (*21*). Kink waves propagate along magnetic field lines, so the inferred wave propagation directions indicate the coronal magnetic field directions projected onto the POS (*22*). Therefore, our global maps represent both the strength and direction of the coronal magnetic field projected onto the POS.

The UCoMP observations provided global coronal magnetograms with an average frequency of approximately once every two days.

**Evolution of the coronal magnetic field**

Previous mapping of the global coronal magnetic field using CoMP (*21, 22*) determined the field strength and direction in the height range of ~1.05 to 1.35 solar radii, and was mostly limited to latitudes ≲50°. The UCoMP maps extend that coverage to higher altitudes, sometimes reaching ~1.6 solar radii (Fig. 1), and to nearly all latitudes, including the polar regions [albeit at lower altitudes and restricted to locations with strong signals (Fig. 1)]. The magnetic field in solar polar regions plays a role in the progression of solar cycles, with the field strength at solar minimum sometimes being used to predict the strength of the following solar cycle (*29*).

From the 114 global maps of coronal magnetic field strength, we constructed *(25)* Carrington maps (maps representing the evolution of solar magnetic field, density or temperature over successive solar rotations; the X-axis of a Carrington map represents the longitude over time, with newer data on the left and older data on the right; the Y-axis represents latitude) of magnetic field strength and plasma density at various heights in the corona during different Carrington rotations of the Sun (e.g., Fig. 2). These maps show the evolution of the coronal magnetic field over eight months, at nearly all latitudes and across different heights as the Sun rotates. We reprojected these Carrington maps for each Carrington rotation onto a spherical coordinate system, from which we extracted the magnetic field distributions in spherical shells with different radii from the solar center (movie S3). Fig. 3 shows three examples of the spherical distributions in two altitude ranges: 1.10 to 1.15 and 1.20 to 1.25 solar radii. For comparison, Figure 3 also shows the spherical distributions of the corresponding radial magnetic field on the photosphere, as measured by the Helioseismic and Magnetic Imager (HMI) (*30*) on the Solar Dynamics Observatory spacecraft. To facilitate comparison, the angular resolution of HMI magnetic field map has been degraded to match that of the UCoMP coronal maps (*25*).

We identified a pattern in the magnetic field structures at different atmospheric layers: regions with strong field in the photosphere generally show strong field in the corresponding coronal measurements (e.g., regions pointed by the white arrows in Fig. 3). This coherence implies that the magnetic fields are connected between different layers of the solar atmosphere. However, we also identified localized patches of strong photospheric field that do not coincide with strong-field features in the corona (e.g., the region pointed by the black arrow in Fig. 3). Possible explanations of this difference include strong expansion of flux tube with height (resulting in a decrease in magnetic field at larger heights due to flux conservation) and low-lying closed magnetic structures (magnetic field lines are closed at lower heights and do not reach higher layers, leading to stronger field at lower heights and weaker field at higher altitudes).

Previous observations of photospheric magnetic fields have shown that newly emerging active regions tend to appear at similar locations as previous active regions (*31, 32*). This behaviour is related to the recurrent emergence of magnetic flux at specific longitudinal sectors, known as active longitudes. The active regions that emerged during our observational period typically did not survive a full solar rotation (see Supplementary Text). The recurrence of strong-field regions at similar longitudes is visible in the coronal Carrington maps (e.g., Fig. 2, B and E, regions marked by the black ellipses), indicating that active longitudes extend their influence as far as the coronal magnetic field.

**Comparison with theoretical models**

Previous work has used three-dimensional MHD models to study the evolution of the coronal magnetic field and its coupling with plasma in the solar corona (*33*). Some models use the observed photospheric magnetic fields to predict coronal emission and magnetic field structures (*33*). For comparison with our LOS-integrated observations, we computed (*25*) coronal models using the Magnetohydrodynamic Algorithm outside a Sphere (MAS) MHD code (*34*) (hereafter MAS models). We generated three MAS models using the photospheric magnetograms during three Carrington rotations (designated CR2254, CR2258 and CR2260). These three were chosen because they correspond to the observations shown in Fig. 1. From the three MAS models we extracted the predicted POS component of the coronal magnetic field strength and direction, the Fe XIII 1074.7 nm line emissivity (the spectral energy released per unit volume per unit time) at each grid point along each LOS within the FOV. From these model data, we calculated the predicted LOS emissivity-weighted physical parameters as:

$$\bar{f} = \frac{\int f_i \cdot \varepsilon_i \, \mathrm{d}l}{\int \varepsilon_i \, \mathrm{d}l} \qquad (2)$$

where $f_i$ is the physical parameter of interest (either the POS component of magnetic field strength $B_{\mathrm{POS}}$ or its direction $\Psi_{\mathrm{POS}}$), $\varepsilon_i$ is the line emissivity, $i$ is an index labelling each location along the LOS, and $\mathrm{d}l$ is the differential distance along the LOS.

The resulting maps of the model-predicted LOS emissivity-weighted $B_{\mathrm{POS}}$ and $\Psi_{\mathrm{POS}}$ are shown in Figure 4, which we compare to the corresponding UCoMP observations in Fig. 1 (see Supplementary Text). Figure 5 shows a quantitative 2D comparison between the models and observations. We found similar features in the observations and model predictions, especially in low- and mid-latitude regions (Fig. 5, A and B). This similarity suggests that the UCoMP measurements of the corona are generally consistent with being LOS emissivity-weighted, as we assumed in our analysis.

We also identified some discrepancies between the MAS models and the observations, which we attribute to the limitations and assumptions in the modelling (*25*). For example, the adopted MAS models have lower spatial resolutions than the observations, so lack some observed small-scale structures (e.g., the fibril-like structures in active regions in the middle column of Fig. 1). The photospheric magnetograms used as boundary conditions for the models were constructed using HMI observations taken over 27 days, not specific to the days of our UCoMP observations. The discrepancies in magnetic field strengths are largest in higher-latitude regions (Fig. 5C), potentially because photospheric magnetic field measurements are less reliable at high latitudes due to projection effects and missing entirely in polar regions. The discrepancies in the coronal magnetic field directions between the models and observations have less of a trend with latitude. We attribute some of the discrepancies to the process of LOS emissivity-weighting of the MAS model output (see Supplementary Text).

A comparison of our observations with the potential field source surface (PFSS) models (*25*) has also been made, revealing some large-scale similarities but also numerous discrepancies. These discrepancies are primarily due to the lack of information on plasma temperature and density, as well as the assumptions inherent in the PFSS models (see Supplementary Text).

## Summary and conclusions

Our results demonstrate routine measurements of the coronal magnetic field over a period of eight months, with an average cadence of once every two days. From these measurements, we generated Carrington maps of the coronal magnetic field at all latitudes, for different altitudes and spanning multiple solar rotations. We found varying field strengths from less than 1 Gauss to ~20 Gauss within 1.05-1.6 solar radii and a sign of active longitude in the coronal magnetic field. Comparison with photospheric observations indicates that some strong-field regions at the photosphere are connected to strong-field regions in the corona and others are not. MHD models of the coronal magnetic field are generally consistent with the observations, but do not match all the observed features.

**Acknowledgments:** We thank Michael Galloy for running the UCoMP data-processing pipeline, the SDO/HMI and SUVI teams for making their data publicly available, Giuliana de Toma for helpful discussions on data processing, and Yuhang Gao for helpful discussions on coronal seismology. HMI is an instrument on SDO, a mission of NASA's Living With a Star Program.

**Funding:** H.T., Z.Y and X.L. are supported by the National Natural Science Foundation of China grant 12425301 and the National Key R&D Program of China (Nos. 2021YFA0718600 & 2021YFA1600500). H.T. also acknowledges support from the New Cornerstone Science Foundation through the Xplorer Prize. R.J.M. is supported by a UKRI Future Leader Fellowship (RiPSAWMR/T019891/1). The UCoMP instrument was developed with support from the National Science Foundation (NSF), and is located at the Mauna Loa Solar Observatory, operated by the High Altitude Observatory of the National Center for Atmospheric Research (NCAR). NCAR is a major facility sponsored by NSF under Cooperative Agreement No. 1852977. Z.Y. acknowledges funding for visits to the High Altitude Observatory from a Newkirk Fellowship awarded by NCAR.

**Author contributions:** H.T. led the project. Z.Y. analyzed the data and generated the figures, tables and movies under the supervision of H.T.. H.T. and Z.Y. wrote and revised the manuscript. S. T. developed the UCoMP instrument, planned the observing sequences and processed the raw data. Z.Y., X.L. and R.M contributed the modified wave-tracking method. S.G. developed the software for the calculation of observables from the MAS models. C.D. contributed to the development of the MAS models. All authors discussed the results and commented on the manuscript.

**Competing interests:** The authors declare no competing interests.

**Data and materials availability:** The UCoMP data can be obtained at https://mlso.hao.ucar.edu/mlso_data_calendar.php?calinst=ucomp. The details of UCoMP data used in this work are listed in Table S1. The SUVI data were obtained from https://www.ngdc.noaa.gov/stp/satellite/goes-r.html, the files we used are specified in the supplementary material. The HMI synoptic magnetograms were obtained from the Joint Science Operations Center http://jsoc.stanford.edu/ajax/lookdata.html?ds=hmi.Synoptic_Mr_720s; we used files named "hmi.Synoptic_Mr.2254.fits", "hmi.Synoptic_Mr.2258.fits", "hmi.Synoptic_Mr.2259.fits", "hmi.Synoptic_Mr.2260.fits", "hmi.Synoptic_Mr.2261.fits" and "hmi.Synoptic_Mr.2262.fits". The source codes, output and scripts of our MAS models are available at Zenodo (*35*). The PFSS model used in this work is available in hierarchical data format at https://www.lmsal.com/solarsoft/archive/ssw/pfss_links_v2/Bfield_20220221_180328.h5. Source codes for the modified wave tracking are available at Zenodo (*36*).

**Supplementary Materials**

Materials and Methods

Supplementary Text

Figures S1-S4

Table S1

References (*37-69*)

Movies S1 to S3

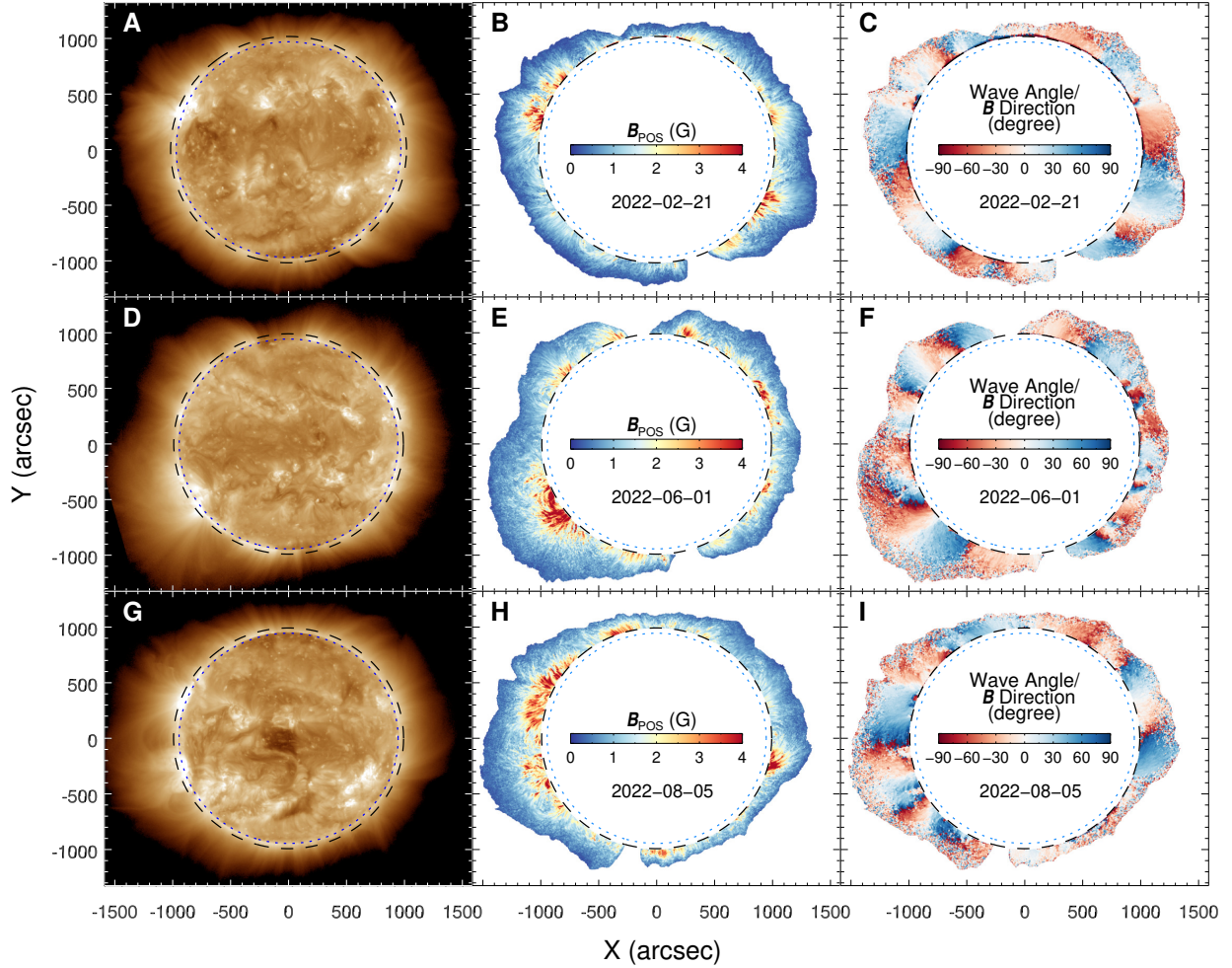

**Fig. 1. Three example global coronal magnetic field maps and corresponding ultraviolet images.** For each panel, X is the east-west direction and Y is the north-south direction. The zero point is the center of the solar disk. (**A**) High-dynamic-range SUVI image of 19.5 nm emission intensity (*25*) from 19:48 to 19:52 Universal Time (UT) on 21 February 2022. (**B**) UCoMP map of the coronal POS magnetic field strength (color bar) on 21 February 2022 (labelled in year-month-day format). A median filter of 3×3 pixels has been applied to reduce noise. (**C**) Same as panel B, but for the magnetic field direction projected onto the POS. Angles are measured with respect to the local radial direction. Positive angles are counterclockwise and negative angles are clockwise. (**D**) Same as panel A but for 20:48 to 20:52 UT on 1 June 2022. (**E** and **F**) Same as panels B & C but for 1 June 2022. (**G**) Same as panel A but for 19:48 to 19:52 UT on 5 August 2022. (**H** and **I**) Same as panels B & C but for 5 August 2022. The uncertainties on the measured magnetic field strength and direction are shown in Figures S2 and S3, respectively. Corresponding maps for all 114 datasets are shown in movie S2. In each panel, the dotted and dashed circles mark the edge of the solar disk (limb) and the inner boundary of the UCoMP FOV, respectively. In all panels of the middle and right columns, the blank parts in the middle are due to the obscuration by the coronagraph occulter, the blank parts in the outer regions are pixels with insufficient S/N.

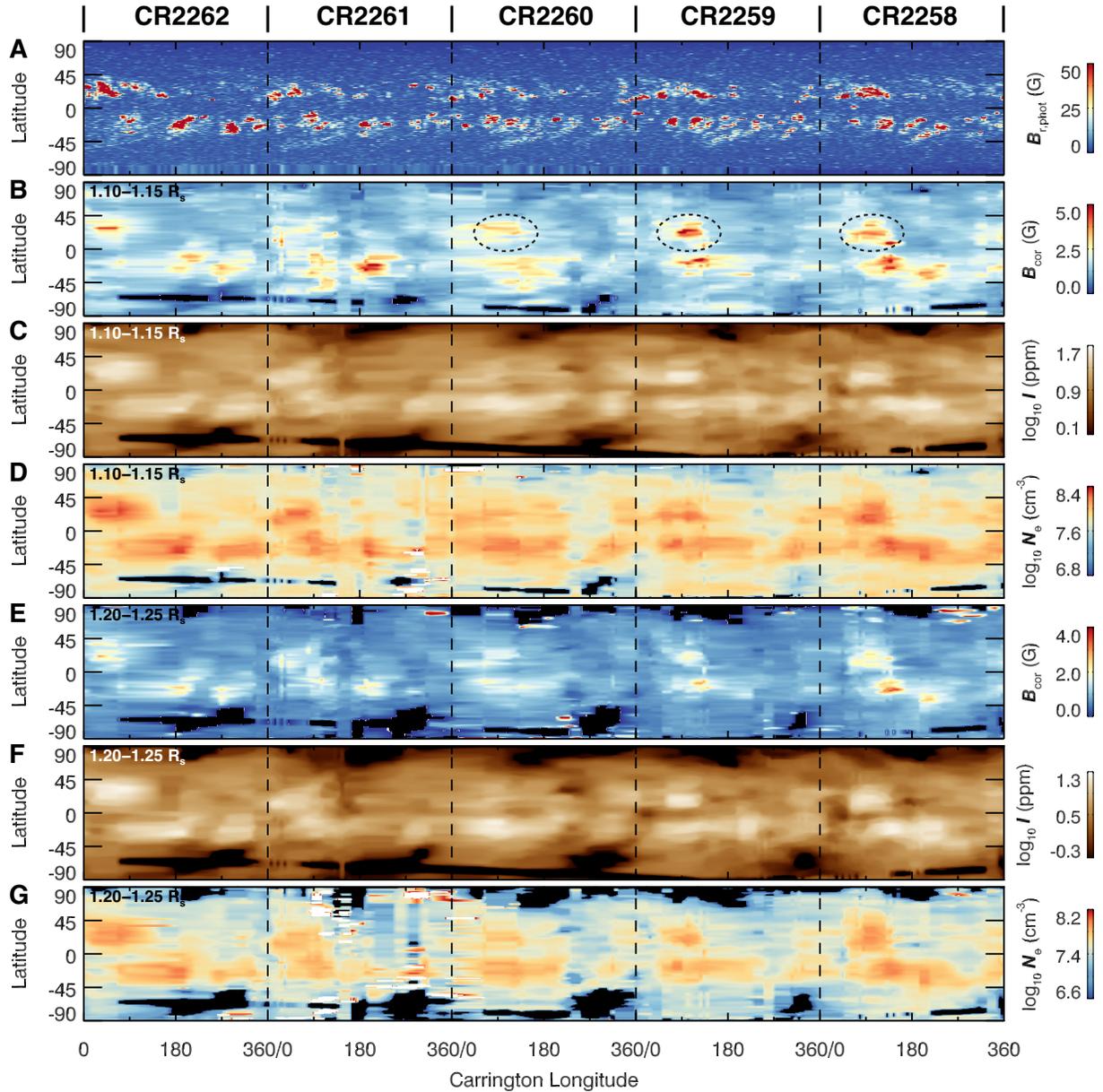

**Fig. 2. Carrington maps of the photospheric and coronal magnetic fields, Fe XIII 1074.7 nm line intensity and electron density.** Here time increases from the right to the left following the convention of Carrington maps. The maps show each quantity (color bars) as functions of solar latitude and Carrington longitude, for five Carrington rotations (CR2258-CR2262) separated by dashed vertical lines. **(A)** The radial component of the photospheric magnetic field strength ($B_{r,phot}$) measured by HMI. **(B-D)** The coronal magnetic field strength ($B_{cor}$), Fe XIII 1074.7 nm line intensity ($I$) and electron density ($N_e$), respectively, each averaged between 1.10 and 1.15 solar radii ($R_S$). **(E-G)** Same as panels B-D but averaged between 1.20 and 1.25 $R_S$. The coronal measurements are non-uniform in longitudes due to data gaps in the 114 datasets, and the corresponding Carrington maps have been interpolated to a consistent angular resolution (1°) along the longitude axis, resulting in elongated features. The black and white areas on the maps represent

regions with low S/N and anomalously high values, respectively. The black ellipses in panel B indicate the recurrence of strong coronal magnetic field at similar longitudes.

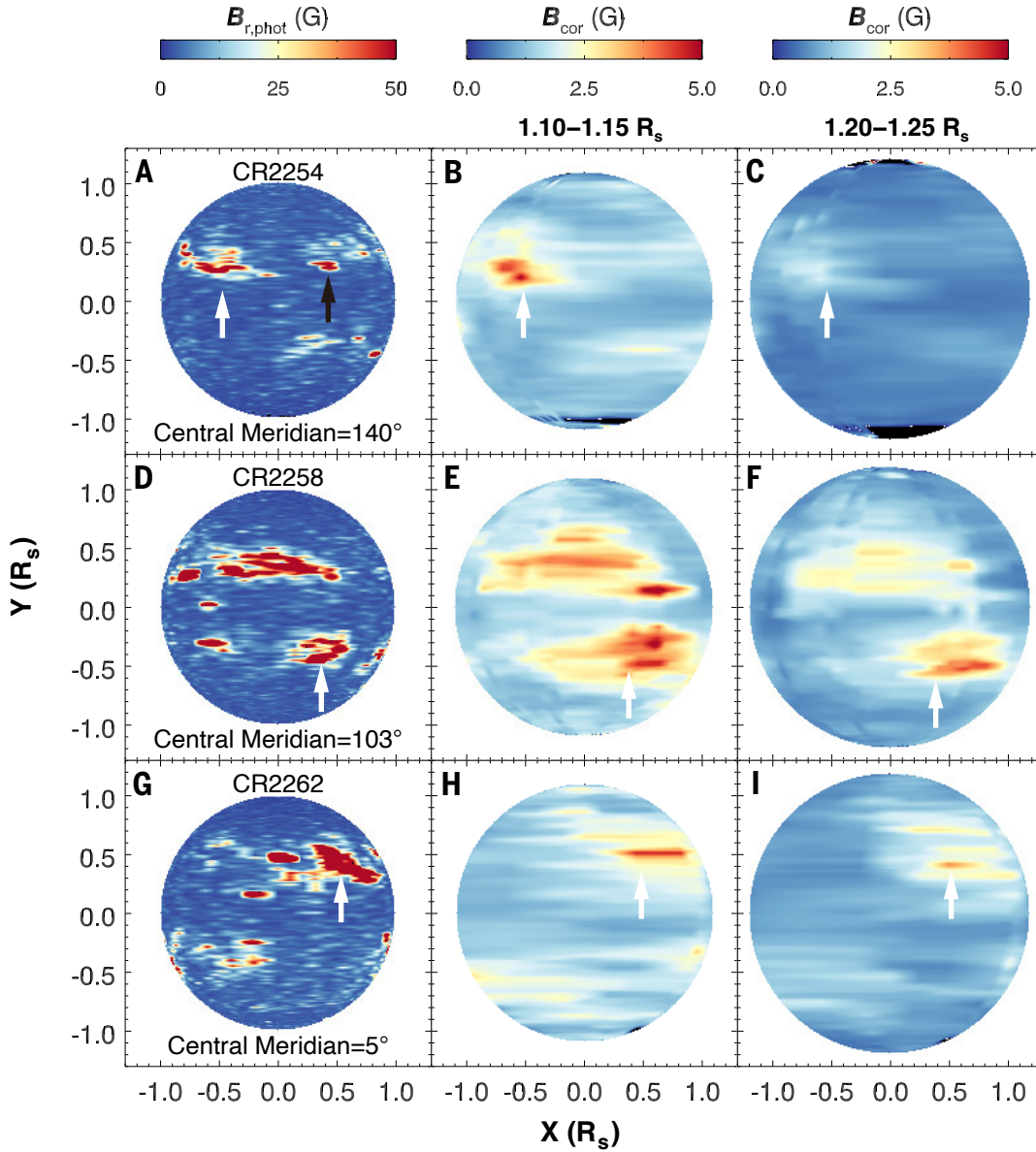

**Fig. 3. Spherical distributions of the magnetic field strength in different atmospheric layers.** (**A**) The radial photospheric magnetic field (color bar) from HMI during CR2254, projected onto a sphere with central meridian at 140°. (**B**) Same as panel A but for the coronal magnetic field from UCoMP averaged between 1.10 and 1.15 solar radii. (**C**) Same as panel B but averaged between 1.20 and 1.25 solar radii. (**D-F**) Same as panels A-C but during CR2258 and projected with a 103° central meridian. (**G-I**) Same as panels A-C but during CR2262 and projected with a 5° central meridian. Movie S3 is an animated version of this figure, showing all meridians over five Carrington rotations. The white arrows indicate several coherent strong-field regions across different layers, whereas the black arrow points to a region with strong field only in the photosphere.

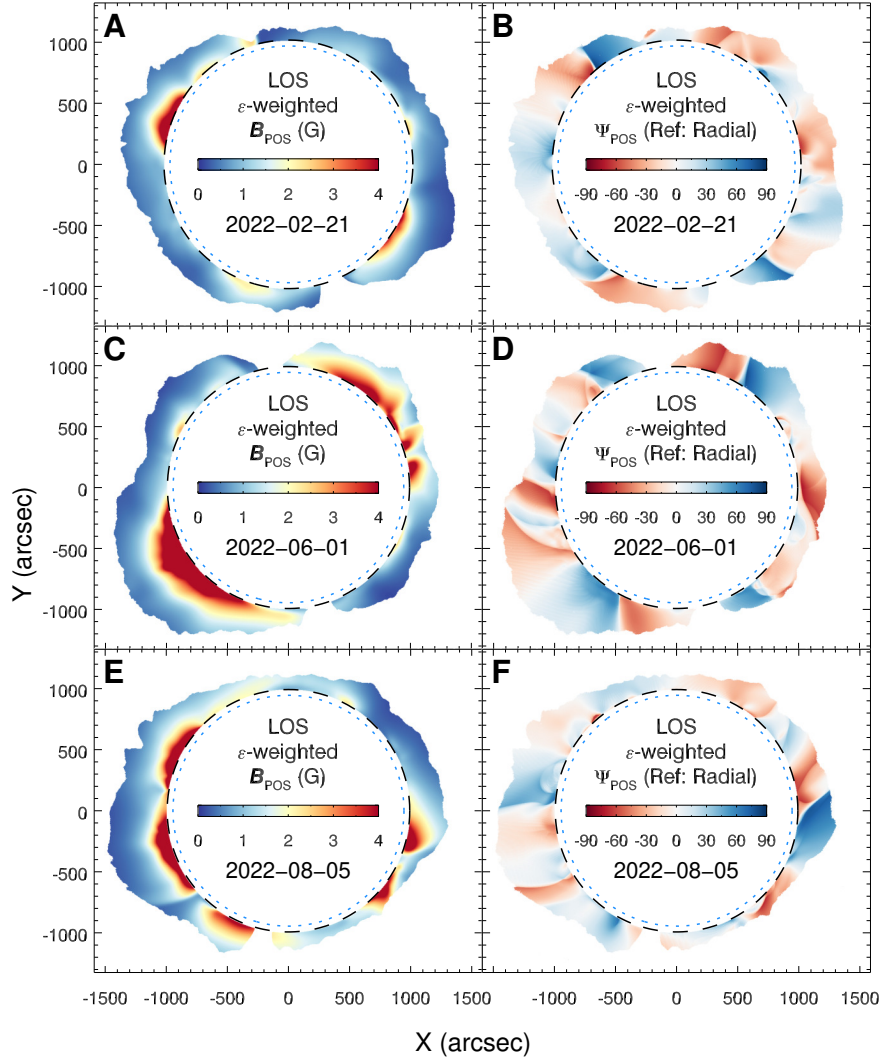

**Fig. 4. Predictions from the MAS models.** Global maps are shown for (**A**) the LOS emissivity-weighted coronal magnetic field strength ($B_{POS}$) and (**B**) magnetic field direction ($\Psi_{POS}$, defined as in Figure 1) predicted by the MAS model of 21 February 2022. Here the label "Ref: Radial" means that the angles are determined with respect to the local radial direction. (**C** and **D**) Same as panels A & B but for the 1 June 2022 model. (**E** and **F**) Same as panels A & B but for the 5 August 2022 model. All these models were constructed using the photospheric magnetic fields observed by HMI as boundary conditions *(25),* and were compared (see Supplementary Text) to the corresponding UCoMP observations (Figure 1). Color bars and circles are the same as in Figure 1.

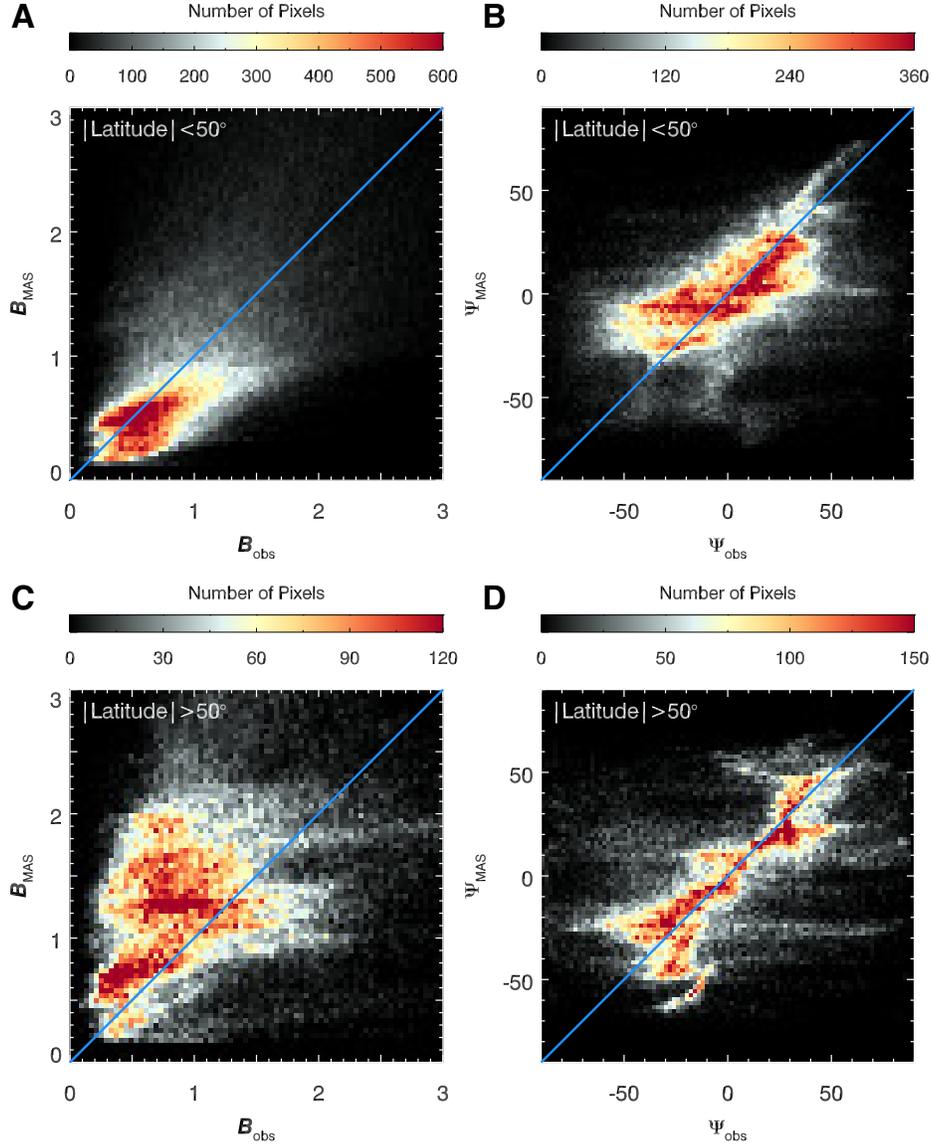

**Fig. 5. Comparison between the MAS models and UCoMP observations.** Two-dimensional histograms compare the measured POS coronal magnetic field (labelled with subscript "obs") to the LOS emissivity-weighted magnetic field predicted by the MAS models (labelled with subscript "MAS"). Each panel shows the number of pixels (color bar) in the maps with each combination of magnetic field strength ($B_{obs}$, $B_{MAS}$) or direction ($\Psi_{obs}$, $\Psi_{MAS}$). The blue diagonal line indicates 1:1 correspondence. **(A)** Comparison between the magnetic field strength in the models and observations for pixels in latitudes between -50° and +50° of the three example maps shown in Figures 1 and 4. **(B)** Same as panel A but for the field direction. **(C)** Same as panel A but for latitudes greater than 50° in both the southern and northern hemispheres. **(D)** Same as panel C but for the field direction.

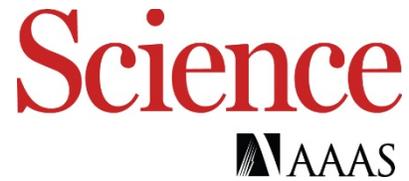

## Supplementary Materials for

**Observing the evolution of the Sun's global coronal magnetic field over eight months**

Zihao Yang, Hui Tian*, Steven Tomczyk, Xianyu Liu, Sarah Gibson, Richard J. Morton, Cooper Downs

*Corresponding author. Email: huitian@pku.edu.cn

**The PDF file includes:**

    Materials and Methods
    Supplementary Text
    Figures S1-S4
    Table S1
    References (*37-69*)

**Other Supplementary Materials for this manuscript:**

    Movies S1 to S3

# 1. Materials and Methods

## 1.1 Observations and Models

### 1.1.1 Upgraded Coronal Multi-channel Polarimeter (UCoMP)

We used data obtained with the UCoMP instrument from 19 February 2022 to 29 October 2022. A total of 114 observations (Table S1) were used to derive the global maps of coronal magnetic field. Compared to CoMP, UCoMP has a larger field-of-view (FOV) ranging from ~1.05 to ~1.6 solar radii (this refers to the FOV that can usually be used for coronal magnetic field diagnostics; the FOV recorded on the detector extends from ~1.05 to 2 solar radii in the east-west direction and from ~1.05 to ~1.6 solar radii in the north-south direction) and a slightly higher spatial resolution of ~6″. UCoMP observes a range of spectral lines which are sensitive to plasma at different temperatures; we used the intensity profiles of the two infrared Fe XIII lines at 1074.7 nm and 1079.8 nm.

For each observation set, a sequence of continuous Fe XIII 1074.7 nm intensity images at three different wavelength positions (1074.59 nm, 1074.70 nm and 1074.81 nm) were used to determine the wave parameters. The data used for wave tracking consists of approximately 120 frames (varies for different datasets, see Table S1) with a time cadence of ~33.5 s. These images were first coaligned. For each time step, we then fitted a Gaussian function to the three-point line profile for every pixel within the FOV, to determine the line parameters: peak intensity, Doppler velocity (relative to the rest wavelength) and line width (*24*). To correct an apparent east-west trend in the Doppler velocity, a polynomial function was fitted to the variation of the mean Doppler velocity in the east-west direction, then subtracted from the values (*24*). We also identified a time-dependent variation in the mean Doppler velocity throughout the continuous observation on each day. To correct for this trend, a linear function was fitted to the time series of the mean Doppler velocity, then also subtracted from the data. Fig. S1 shows an example of the two corrections.

An additional set of nearly simultaneous Fe XIII 1074.7 nm and 1079.8 nm intensity images were used to perform density diagnostics. This series of intensity data was observed several tens of minutes before the wave observations. For each spectral line, multiple frames (typically 4, varying for different datasets, see Table S1) of intensity images at each wavelength position (for the Fe XIII 1074.7 nm line: 1074.59 nm, 1074.70 nm and 1074.81 nm; for the Fe XIII 1079.8 nm line: 1079.69 nm, 1079.80 nm and 1079.91 nm) were coaligned and averaged to enhance the signal-to-noise ratio (S/N). These images were also coaligned to the Doppler velocity images.

We only performed analysis for coronal regions with sufficient S/N, determined as regions where the Fe XIII 1074.7 nm intensity exceeds 1.8 parts per million (ppm) of the intensity on the solar disk.

### 1.1.2 Solar UltraViolet Imager (SUVI)

We used archival extreme-ultraviolet (EUV) imaging observations from the SUVI instrument on the GOES spacecraft. SUVI provides coronal EUV images with a FOV that matches the FOV of UCoMP.

Figure 1 shows the SUVI level-2 composite images from the GOES archive, which are high-dynamic-range images generated from multiple exposures with short and long integration times. These composite images were observed using the 19.5 nm channel of SUVI with a spatial pixel size of 2.5″. Fig. 1, A, D and G used observations taken from 19:48 UT to 19:52 UT on 21 February 2022, from 20:48 UT to 20:52 UT on 1 June 2022 and from 19:48 UT to 19:52 UT on 5 August 2022, respectively.

### 1.1.3 Helioseismic and Magnetic Imager (HMI)

To compare with our synoptic coronal magnetic field maps (Carrington maps of the coronal magnetic field), we used archival HMI synoptic photospheric magnetograms (radial component) spanning several Carrington rotations from CR2254 to CR2263. Each synoptic magnetogram was constructed from HMI 720-s line-of-sight full-disk magnetograms. Initially, the 720-s line-of-sight magnetograms were converted to radial magnetograms. Then the magnetograms were interpolated onto a Carrington coordinate system. For each Carrington longitude, the magnetic field values were obtained by averaging the values from magnetograms with central meridians in close proximity to the corresponding Carrington longitude. The final product of the synoptic magnetogram combines HMI 720-s disk magnetograms over a full solar rotation (~27.27 days) and has original dimensions of 3600 pixels × 1440 pixels. For the three examples in Fig. 4, we used HMI synoptic magnetograms from CR2254, CR2258 and CR2260 as the boundary condition for coronal MHD models. A boundary magnetogram has approximately 300×144 mesh points in the $\phi - \theta$ directions. Here $\phi$ and $\theta$ refer to the azimuthal and polar angles, respectively. A uniform grid in the $\phi$ direction and a slightly non-uniform grid in the $\theta$ direction were employed. The resolution in the $\theta$ direction is matched to that in $\phi$ at the equator, and is slightly coarser towards higher latitudes. The pixel size at the equator of the boundary magnetogram is approximately $0.021\,R_s \times 0.021\,R_s$.

### 1.1.4 Magnetohydrodynamic model

For comparison with the observations, we used the Magnetohydrodynamic Algorithm outside a Sphere (MAS) model, from Predictive Science, which is a 3D MHD model of the thermodynamics and magnetic structure of the solar corona (*34*). A detailed description of the MAS model setup has been published previously (*37*). The MAS model solves resistive MHD equations including thermal conduction, radiative loss and coronal heating terms. The boundary condition for the model is the synoptic photospheric magnetogram (radial component) derived from HMI observations. We used MAS models generated using the synoptic photospheric magnetograms from CR2254, CR2258 and CR2260 as the boundary conditions.

The models have dimensions of 255×144×300 ($r, \theta, \phi$), with the radial distance ($r$) ranging from 1 to 30 solar radii. The model grid is uniform in the $\phi$ direction and slightly non-uniform in the $\theta$ direction. The voxel size in the $\phi - \theta$ directions at the equator is approximately $0.021\,r \times 0.021\,r$, where $r$ is the distance from solar center. We adopted the empirical thermodynamic heating model 2 in the MAS code, similar to the one used in previous work (*34, the third heating model*). This heating model accounts for different heating functions with varying heating rates for active regions, quiet-Sun regions and coronal holes (*34, 38*).

Previous studies often applied a factor of 1.4 to correct the radial component of the photospheric magnetic field from HMI measurements before using them as input for MAS models. Those studies included this factor to account for differences between measurements from HMI and the Michelson Doppler Imager (MDI) (*39*) on the Solar and Heliospheric Observatory (*40*), and was used for MAS models calculated while MDI was operating (*33*). To maintain consistency with previous works, we also multiplied the HMI photospheric magnetograms by a factor of 1.4 before running the MAS models for the three Carrington rotations.

The MAS model output provides three-dimensional distributions of the coronal magnetic field vector, density and temperature in a spherical coordinate system. From these model outputs, we calculated the coronal emissivity and POS component of the magnetic field strength and direction along each line-of-sight (LOS) using Equation 2. The results are shown in Figure 4.

### 1.1.5 Potential Field Source Surface (PFSS) model

We also used the PFSS model for comparison. This is also a coronal magnetic field model using the synoptic photospheric magnetogram as the boundary condition. It extrapolates the magnetic field into the corona under the assumption of a potential field, containing no current or magnetic free energy. We used a PFSS model extrapolated from the synoptic photospheric magnetogram sampled at 18:03:28 UT on 21 February, 2022.

### 1.1.6 Atomic data

We adopted atomic data from the CHIANTI database version 10.0 (*41, 42*). For our calculation, the spontaneous radiative decay rates and the electron collisional rates of Fe XIII ions (*43*) implemented in the CHIANTI database were used.

### 1.1.7 Synthetic observables

We used the FORWARD software package implemented within the SOLARSOFT framework (*44*). Using FORWARD, we obtained the POS component of the coronal magnetic field along each LOS from the MAS models and investigated the LOS integration effect. We also synthesized the emissivity of the Fe XIII 1074.7 nm line along each LOS using FORWARD. The model magnetic field and the synthesized emissivity were combined to derive the LOS emissivity-weighted coronal magnetic field strength and direction using Equation 2. The POS component of the coronal magnetic field on the POS crossing solar center and perpendicular to the LOS was also obtained from the PFSS model using FORWARD.

## 1.2 Methods

### 1.2.1 Derivation of the wave phase speed

Previous studies showed that the prevalent transverse waves observed with CoMP and UCoMP are kink waves (*17, 19, 21*). In the corona, where the lengths of magnetic flux tubes are generally much larger than the radii of their cross sections (known as the long-wavelength limit or thin flux-tube approximation), the phase speed of kink waves in cylindrical flux tubes can be approximated as

$$v_k \approx \sqrt{\frac{B_n^2 + B_x^2}{\mu_0 (\rho_n + \rho_x)}} \qquad \text{(S1)}$$

which relates the kink speed in a coronal flux tube to the magnetic field ($B_{n,x}$) and density ($\rho_{n,x}$) in the interior (denoted by subscript n) and exterior (denoted by subscript x) of the flux tube. At the coronal heights observed by UCoMP, the coronal plasma-$\beta$ (the ratio between gas pressure and magnetic pressure) is typically $\ll 1$. In such an environment, the pressure balance between the interior and exterior of a flux tube requires $B_n \sim B_x$ (*17, 45, 46*). In addition, the widths of coronal flux tubes (*47*) are usually much smaller than the spatial resolution (~6″) of UCoMP, so we assumed that the observations do not resolve individual flux tubes.

The observed emission in each pixel originates from both regions inside and outside flux tubes. Because the observed infrared line emission is contributed by both collisional excitation (line intensity proportional to the square of electron density $N_e^2$) and photo-excitation (line intensity proportional to the electron density $N_e$), the total line intensity is proportional to $N_e^\alpha$ where $\alpha \in (1, 2)$. Previous model calculations found that the intensity of Fe XIII 1074.7 nm follows a height dependence very close to α=1, indicating that photo-excitation dominates (*48*). We therefore interpreted the derived density at each pixel as the average density inside and outside flux tubes within that pixel. Under such coronal and instrumental conditions, Equation S1 simplifies further to Equation 1. Although Equation 1 is similar in form to the Alfvén speed equation, it describes the properties of kink waves observed by UCoMP. In extreme cases where the derived density is predominantly influenced by higher-density regions (e.g., inside coronal loops) due to higher contributions from collisional excitation, the density contrast between the interior and exterior of the loops could impact the derived magnetic field strengths (*49*). Typical density contrasts of quiescent coronal loops are ~3 to 10 (*50*), so even in the extreme case of the derived density representing the density within coronal loops, the resulting magnetic field strength is biased by ≲ 30%. Previous forward modeling studies of propagating kink waves have demonstrated that Equation 1 has errors <20% for density contrasts of 2 to 15 (*27*).

### 1.2.2 Density diagnostics

We used the intensity ratio between the Fe XIII 1074.7 nm and 1079.8 nm lines as a diagnostic of the electron density in the solar corona (*51, 52*). We used the atomic data from CHIANTI to calculate the theoretical relationship between the line ratio and electron density, and combined this relationship with the observed line ratio to diagnose the electron density at each pixel within the FOV of UCoMP. For further details regarding our method, we refer to the previous publications (*21, 22*).

### 1.2.3 Wave tracking

We modified a wave tracking technique developed in previous work (*17, 21, 22, 26*) to enhance the speed and accuracy of computations.

The wave tracking technique comprises two main steps: the calculation of wave propagation direction and the calculation of wave phase speed. To determine the wave propagation direction, we first obtained a filtered Doppler velocity time series with a regular cadence of ~33.5 s. The power spectrum of the Doppler velocity of these wave observations has a peak frequency of approximately 3.5 mHz, corresponding to a period of around 5 minutes (*18, 53*). Our sampling cadence is 33.5 s, meaning that we have approximately 9 samples within each wave period, sufficient to determine the wave properties. For each pixel within the FOV, we selected a box of 41×41 pixels centered around the pixel. Using cross correlation, we calculated the coherence between the Doppler velocity time series at the target pixel and those at its surrounding pixels. Regions with high coherence have elongated shapes. To calculate the propagation direction, a linear function was fitted to the elongated region. We modified this calculation to use weighted linear fitting, with higher weights assigned to pixels with higher coherence values, because the propagation of transverse waves favors regions with higher coherence. By repeating this process for each pixel in the observed FOV, we obtained a map of wave propagation direction. Fig. 1, C, F and I provide three examples of coronal wave propagation direction maps.

From the wave propagation direction maps, the wave path was derived for each pixel within the FOV. We then constructed a Doppler velocity time-distance diagram for each wave path. We modified this step to address potential issues with averaging and smoothing the time series that could bias the results. In the previous method, a high-S/N velocity time series was first derived as the reference time series from the time-distance diagram by cross-correlating and averaging the time series along the wave path (*17, 26*). However, this averaging process sometimes introduces smoothing artifacts that bias the results. We therefore modified the method by selecting only the time series from the center pixel of the wave path as our reference time series. By cross-correlating the reference times series (the central time series) with other time series along the wave path, we obtained the time lag (number of pixels required to slide one time series over the other to achieve maximum correlation) and the corresponding correlation coefficient after sliding the time series. Weighted linear fitting was applied to the relative position along the wave path versus the time lag. Higher weights were assigned to points with higher correlation coefficients obtained during the cross-correlation analysis. The phase speed at each pixel was then calculated using the weighted linear fits.

Wave reflection and nonlinearity could also impact the wave tracking results. Waves in the corona can undergo reflection during propagation. Transverse waves observed in the corona using CoMP sometimes experienced reflection, resulting in inward and outward propagating components (*26*). We followed previous work (*21, 22, 26*) by isolating the outward propagating component for the calculation of wave propagation direction and phase speed. Compared to the downward propagating component, this component typically exhibits a higher wave power and, consequently, a higher S/N, making it more suitable for wave tracking analysis.

Previous studies have shown that nonlinearity could enhance wave damping (*54, 55*), resulting in reduced wave amplitudes. The calculation of wave phase speed relies on cross-correlation of the Doppler velocity time series, which requires identification of periodic variation in wave amplitude. If the wave amplitude becomes too small to resolve its periodic variation, the accuracy of cross-correlation (as well as the calculation of phase speed) will be compromised. We investigated our observations for signs of wave damping but did not find any, so assumed that its impact is negligible. Previous studies using CoMP observations also found weak damping for propagating kink waves (*56*).

The wave tracking results could also be impacted by loop structures (e.g., loop curvature) and magnetic field nonuniformity (e.g., multi-stranded loops and multiple loops within a pixel) (*27, 57-59*). Previous work (*60, 61*) has found that the loop curvature has a very weak impact on the kink oscillations, so the dispersion relation we derived based on a straight cylindrical flux tube is barely affected. Multi-stranded loops with non-twisting threads often show collective behavior and share the same oscillation pattern (*58, 62*), so our results will not be affected. Previous forward modeling has also demonstrated that multi-stranded loops have minimal impact on the validity of Equation 1 in deriving the magnetic field strength (*27*). In the presence of separate loops with different oscillation properties, the interaction between loops has minimal impact on the wave properties if the loops are well-separated (*63, 64*). However, in regions where different loops are in close proximity in distance (within approximately the loop diameter), the derived wave parameters could be affected (*63, 64*), which would lead to higher uncertainty in our measurements.

### 1.2.4 Estimation of uncertainties

Equation 1 shows that the uncertainty in the magnetic field strength can be obtained by propagating the uncertainties in the phase speed and the density.

During the wave tracking process, the phase speed was determined through weighted linear fitting. Therefore, we determined the uncertainty in the phase speed using the uncertainty in the fitted slope (*21*).

Previous work (*21, 22*) has shown that the uncertainty in density is associated with the uncertainties of the measured line intensities and the inherent systematic uncertainty in the theoretical density-line ratio relationship. Following those studies, we calculated the uncertainty introduced by the theoretical relationship. The UCoMP measurement uncertainty was determined using the same approach in previous work (*21, 22*). To convert the uncertainty in ppm to photon counts, a conversion factor $k$ was used. UCoMP has a different $k$ to CoMP, which was determined using the values of parameters "FLATDN" and "BOPAL" from the headers of the level-1 .fits files ($k = \frac{\text{FLATDN}}{\text{BOPAL}} * 4.12$).

Fig. S2 shows the estimated uncertainties for the inferred coronal magnetic field strengths shown in Fig. 1. The uncertainty is <20% at most locations.

As discussed in Sect. 1.2.3, the magnetic field direction (wave propagation angle) was determined by using linear fitting that minimizes the perpendicular distances of the data points in the high-coherence region from the line being fitted (*26, 65*). The parameter to be minimized, $R_{\perp 0}^2$, is

$$R_{\perp 0}^2 = \sum_{i=1}^{N} \frac{(Y_i - KX_i)^2}{1+K^2} \quad \text{(S2)}$$

where $X_i$ and $Y_i$ are the coordinates of each data point in the high-coherence region, $K$ is the slope of the fitted line, $N$ is the total number of data points, and $i$ is the index of data points. To estimate the uncertainty associated with the fitted angle, a small perturbation of $\Delta\alpha = 2°$ was applied to the best-fitted angle $\alpha_0$ (related to the slope of the best-fitted line, $K_0$), resulting in $\alpha_- = \alpha_0 - \Delta\alpha$ and $\alpha_+ = \alpha_0 + \Delta\alpha$. The angles $\alpha_-$ and $\alpha_+$ correspond to slopes $K_-$ and $K_+$, respectively. The to-be-minimized parameter, $R_\perp^2$, increases as the line slope $K$ deviates from $K_0$ towards $K_-$ and $K_+$:

$$R_{\perp,-}^2 = \sum_{i=1}^{N} \frac{(Y_i - KX_i)^2}{1+K^2} \text{ and } R_{\perp,+}^2 = \sum_{i=1}^{N} \frac{(Y_i - KX_i)^2}{1+K_+^2} \quad (S3)$$

If the data points are more closely aligned with the fitted line, any deviation in the slope causes the data points to move away from the deviated line. Therefore, even a slight change in the slope can result in a large variation in $R_\perp^2$, indicating that the fitted slope is more precisely determined (the uncertainty in the fitted slope is small). Conversely, when the data points are more scattered, a small change in the fitted slope causes only a small variation in $R_\perp^2$. This indicates that the linear fit is poorly constrained, resulting in a larger uncertainty in the fitted slope. The uncertainty in the fitted angle $\sigma_\alpha$ was estimated to be inversely related to the variation in $R_\perp^2$ in the presence of a small perturbation of the fitted angle:

$$\sigma_\alpha = \frac{\Delta\alpha}{\sqrt{(R_{\perp,-}^2 - R_{\perp 0}^2) + (R_{\perp,+}^2 - R_{\perp 0}^2)}} \quad (S4)$$

This equation reflects the uncertainty in the angle calculation due to the scatter of data points and the fitting process.

Fig. S3 shows the estimated uncertainties for the inferred coronal magnetic field directions (wave propagation angles) shown in Fig. 1. We found that the uncertainties are generally <2 degrees at most locations. However, in higher coronal regions with a low S/N, the uncertainties are larger.

### 1.2.5 Construction of the coronal Carrington maps

From the UCoMP observations, we derived 114 maps of global coronal magnetic field. Each map consists of the east limb and west limb, which correspond to different Carrington longitudes. Assuming minimal coronal structure evolution during one solar rotation, the measuremens from west and east limbs represent coronal magnetic fields approximately 1/4 Carrington rotation before and after the time of observation, respectively.

For each limb measurement on each map, we calculated the latitudinal distribution of coronal magnetic field strength averaged over a distance of 0.01 solar radius. This range extends from 1.05 solar radii to the outermost height, which varies between datasets. The spatial size of the angular resolution element changes with height (solar radii), which we took into account during the averaging. By associating each map with its central meridian and corresponding Carrington rotation number, we calculated the Carrington longitudes of the two latitudinal distribution stripes. The 114 maps yielded 228 stripes of latitudinal distributions corresponding to different Carrington longitudes across various Carrington rotations.

To construct the synoptic Carrington maps, we interpolated the latitudinal distribution stripes along the Carrington longitude direction, to obtain a Carrington map with its Y-axis representing latitude and its X-axis representing Carrington longitude. We derived Carrington maps of coronal magnetic field strength, intensity and density across various coronal heights. Although the HMI synoptic photospheric magnetograms have been corrected for solar B-angle (the angle between the solar equatorial plane and the ecliptic plane), the latitude in our coronal Carrington maps was not corrected for B-angle. This is because the off-limb measurements are LOS integrated, complicating any correction for B-angles. Our data span B-angles between only -7.23 and +7.23 degrees, so we assumed that the effect is negligible.

For the Carrington maps of photospheric magnetic field, we used the synoptic maps from HMI measurements during CR2254 to CR2263. To match the angular resolution of the coronal Carrington maps, we degraded the angular resolution of the photospheric Carrington maps as follows: i) Converted the Y-axis of the original HMI photospheric Carrington map from sine latitude to colatitude (90° minus latitude), then interpolated the Y-axis to uniform colatitudes $\theta_{co}$; ii) Took the absolute value of the radial component of the magnetic field $|B_r|$, and multiplied by $\sin \theta_{co}$ to ensure that flux is preserved during degrading; iii) Degraded the Carrington map to the desired angular resolution by dividing the map into small regions and averaging values inside each small region.

Fig. 2 shows examples of Carrington maps of photospheric magnetic field measured by HMI, coronal magnetic field strength, line emission intensity and electron density averaged between 1.10 and 1.15 solar radii, and between 1.20 and 1.25 solar radii, derived from the UCoMP observations. These Carrington maps cover five Carrington rotations from CR2258 to CR2262. We found that regions with higher density generally exhibit higher coronal magnetic field strengths, which is expected since stronger-field regions such as active regions generally have higher densities.

Fig. 3 shows snapshots from the spherical distributions of magnetic field at different heights, obtained by reprojecting the Carrington maps from each Carrington rotation onto a spherical coordinate system.

### 1.2.6   Calculation of LOS emissivity-weighted parameters

To facilitate comparison between the models and observations, investigate any discrepancies between them, and determine the potential effects of LOS integration, we calculated the LOS emissivity-weighted coronal parameters from the MAS models. These LOS weighted parameters were determined using Equation 2. For each pixel within the UCoMP FOV, we computed the LOS distributions of coronal emissivity, $B_y$ (the coronal magnetic field component along X-direction on the POS) and $B_z$ (the coronal magnetic field component along Y-direction on the POS) from one solar radius behind to one solar radius in front of the POS crossing the solar center with an interval of 0.01 solar radius. We followed conventional notations $B_y$ and $B_z$ to describe the two magnetic field components in the models.

The POS component of coronal magnetic field strength is $B_{POS} = \sqrt{B_y^2 + B_z^2}$. We calculated the LOS emissivity-weighted coronal magnetic field strength (POS component) $\overline{B_{POS}}$ for each pixel within the FOV as:

$$\overline{B_{POS}} = \frac{\int_{-1R_s}^{+1R_s} B_{POS,i} \cdot \varepsilon_i \, dl}{\int_{-1R_s}^{+1R_s} \varepsilon_i \, dl} \quad (S5)$$

where $B_{POS,i}$ and $\varepsilon_i$ are the POS component of coronal magnetic field strength and emissivity of Fe XIII 1074.7 nm at location $i$ along the LOS, respectively. The integrals were taken along the LOS from $-1R_s$ to $+1R_s$, because the emissivity drops quickly when it approaches larger heights above the solar surface. We do not perform weighted averaging of magnetic field vectors, but of $B_{POS}$. This is partially connected to the wave observations and the technique of coronal seismology. The magnetic field strength measured using UCoMP wave observations is based on the wave phase speed, which is a scalar quantity that is only dependent on the strength of the magnetic field and not its direction.

Similarly, we computed the LOS emissivity-weighted coronal magnetic field direction (POS component) $\overline{\Psi_{POS}}$. Because the waves propagate along field lines, the measured wave propagation directions provide information about the field directions. The wave propagation directions are solely determined by the field directions. Therefore, we calculated the emissivity-weighted directions without involving magnetic field vectors. Firstly, for each location $i$ along each LOS, we calculated the coronal magnetic field direction (POS component) as

$$\Psi_{POS} = \tan^{-1}\left(\frac{B_z}{B_y}\right) \quad (S6)$$

The resulting value of $\Psi_{POS}$ was then converted to $\Psi_{POS}^{rad}$, the direction with respect to the local radial direction. $\Psi_{POS}^{rad}$ is defined as positive if it is counterclockwise from the local radial direction, and negative otherwise. We limit its range to $(-90°, 90°)$, due to the 180°-ambiguity. The LOS emissivity-weighted coronal magnetic field direction (POS component) $\overline{\Psi_{POS}}$ for each pixel within the FOV was calculated as:

$$\overline{\Psi_{POS}} = \frac{\int_{-1R_s}^{+1R_s} \Psi_{POS,i}^{rad} \cdot \varepsilon_i \, dl}{\int_{-1R_s}^{+1R_s} \varepsilon_i \, dl} \quad (S7)$$

where $\Psi_{POS,i}^{rad}$ and $\varepsilon_i$ are the POS component of coronal magnetic field direction (with respect to local radial direction) and emissivity of Fe XIII 1074.7 nm at location $i$ along the LOS, respectively. The integrals were taken along the LOS from $-1R_s$ to $+1R_s$.

## 2. Supplementary Text

### 2.1 Evolution of coronal magnetic field

Previous photospheric observations have shown a phenomenon referred to as activity nests, in which newly-emerging active regions tend to appear at similar locations as the previous active regions or in close proximity to existing ones (*31, 32*). This behavior is closely associated with the recurrent emergence of magnetic flux at certain longitudinal sectors from beneath the photosphere, known as active longitudes. From photospheric observations of the Sun's front side by SDO/HMI and helioseismic estimations of the far-side photospheric magnetic features (*66-68*; here we used data from June 2022 to August 2022), we found that active regions in our eight-month observational period typically did not survive a full solar rotation; instead, new active regions often emerged in the vicinity of previously existing active regions. For example, the active regions designated AR13030, AR13031 and AR13032 disappeared within one solar rotation, while new active regions AR13053 and AR13055 emerged near the previously dissipated active regions. From our Carrington maps of the coronal magnetic field, we found a recurrent pattern of strong-field features at similar longitudes in the corona (Fig. 2, B and E), which corresponds to the repeated emergence of different active regions in similar longitudinal sectors during different

Carrington rotations. This indicates that active longitudes extend their influence to the coronal magnetic field.

The coronal magnetic field strengths of active regions in the Carrington maps also exhibit fluctuations over several solar rotations. For example, the maximum values of the magnetic field strength between 1.10 and 1.15 solar radii varies from ~10 G to ~20 G in the corona above different active regions during our eight-month observational period. In contrast, the maximum magnetic field strengths in most other regions within this height range are <5 G.

## 2.2 Comparison with MAS models

Coronal observations using optically thin spectral lines are subject to the LOS integration effect. Our analysis assumed that the observed quantities represent the emissivity-weighted results along the LOS. To verify this assumption and investigate the impact of LOS integration on the measured physical quantities, we compared our measurements with predictions from the MAS models.

We calculated the LOS emissivity-weighted coronal magnetic field strength and direction (POS component) from the MAS models as described in Section 1.2.6. Figure 5 compares the coronal magnetic field from the UCoMP observations with the LOS emissivity-weighted quantities from the MAS models. Fig. 5A-B shows that the observational measurements and model predictions are correlated. For Fig. 5A, the Pearson correlation coefficient is 0.64. This supports our assumptions that the measured coronal magnetic field strength corresponds to the LOS emissivity-weighted values, and that the wave propagation direction used to infer the field direction from the observation reflects the coronal magnetic field direction.

Despite the general correlation, we also found discrepancies between the observational measurements and the model predictions. In Fig. 5A the magnetic field strengths in active regions (identifiable as those with stronger observed field) are under-predicted by the models. We attribute this discrepancy to two effects: i) The synoptic photospheric magnetic field maps used as input to the MAS models do not include any time variation or 3D vector field information, and the need for 27 days of data means that they do not include any newly developed active regions or other magnetic structures near the solar limbs. ii) Active regions contain more complex magnetic and thermodynamic structures than quiet-Sun regions, so the LOS integration effect is also more complex and modelling the complicated plasma structures in these regions is more challenging. The MAS models we used have low spatial resolution and adopted empirical heating terms. This could partly explain the absence of fine structures in the model predictions. We noticed that higher-resolution simulations have produced filamentary structures of the magnetic field in LOS emissivity-weighted model predictions (*14*).

We found even larger discrepancies in the magnetic field strength for regions at higher latitudes. Fig. 5C shows the relationship between $B_{\mathrm{MAS}}$ and $B_{\mathrm{obs}}$ for regions with latitudes exceeding 50° in both the northern and southern hemispheres. There is a much larger discrepancy than at lower latitudes: the Pearson correlation coefficient is only 0.34. We attribute the lower correlation in high-latitude regions to the less reliable measurements of the high-latitude photospheric magnetic field, due to projection effects and unobserved polar regions (*69*). There is no trend in Fig. 5C in the discrepancy as a function of coronal magnetic field strength. However, in some regions with lower intensity, large discrepancies appear in the field directions. We attribute this to the higher uncertainty in field direction in such regions due to the lower S/N (Fig. S3).

Some of the discrepancies between observations and model results could instead arise from the calculation of the LOS emissivity-weighted parameters. Although that calculation is a linear operation, the derivations of the phase speed, density, magnetic field strength and direction are all non-linear operations. Therefore, the linear averages of the field strength and direction from the models are only first-order approximations, which could deviate from the observed parameters. Nevertheless, we found that the LOS emissivity-weighted parameters predicted by the models are sufficient for comparison purposes.

We conclude that the observationally measured coronal magnetic fields are generally consistent with the predictions of the models (maps of LOS emissivity-weighted field). We do not expect a perfect correspondence due to the limitations of LOS emissivity-weighting, the absence of real-time information and polar observations in the boundary photospheric magnetograms.

## 2.3 Comparison between observations and PFSS models

We also compared the observations to simpler coronal magnetic field extrapolation models using the potential-field source surface (PFSS) method.

The wave phase speeds we measured from the observations are the projection of the true speeds onto the POS, so the measured field strengths should correspond to the POS projection of the coronal magnetic field (*21, 22*). Due to the LOS integration effect, for each pixel within the FOV the measured magnetic field (both strength and direction) represents the emissivity-weighted average of the POS projection of the magnetic field vector at each grid point along the LOS. Because PFSS models lack information about the plasma temperature and density, we cannot use them to calculate emissivity or emissivity-weighted parameters. We compared the UCoMP measurements with the POS component of the magnetic field on the plane crossing the solar center and perpendicular to the LOS predicted by the PFSS models described in section 1.1.5 above.

Figure S4A-B shows the maps of coronal magnetic field strength from the UCoMP observations on 21 February 2022 and the corresponding PFSS model. To compare the height distributions of the magnetic field, we selected four annulus sectors (corresponding to two active regions and two quiet-Sun regions, labelled in Fig. S4A-B) and plotted the variation of average field strength as a function of radial distance within each sector (Fig. S4C-F).

We found some large-scale similarities between observational measurements of UCoMP and the predictions from the PFSS model. For example, both maps show higher coronal magnetic field strengths above active regions. However, we also identified many discrepancies. For example, the PFSS model predicts high field strengths in polar regions, which could be due to the unreliable magnetic field information of these regions in the boundary photospheric magnetograms. The average coronal magnetic field strengths from the observations and models follow similar decreasing trends with height, but differ in magnitude. As discussed in previous work (*21*), such discrepancies could arise from the violation of the potential-field assumption (made by the PFSS models) in certain regions, particularly above active regions where the coronal magnetic field is unlikely to be potential field. We also used a different definition of $B_{\text{POS}}$ between the observational measurement and the PFSS model prediction. For the observational measurement, the magnetic field was determined as the emissivity-weighted average of the POS projection of the magnetic field vector at each grid point along the LOS, whereas the PFSS model provides only the magnetic field on a single POS. This difference should also contribute to the discrepancies between the two results. In addition, the various factors that can potentially impact the accuracy of our method, as discussed in Sect. 1.2.3, may also contribute to the discrepancies found here. The comparisons with both MAS models and PFSS models suggest that coronal MHD models align more closely with observational data, and are more suitable for an investigation of the impact of LOS integration effect on coronal observations.

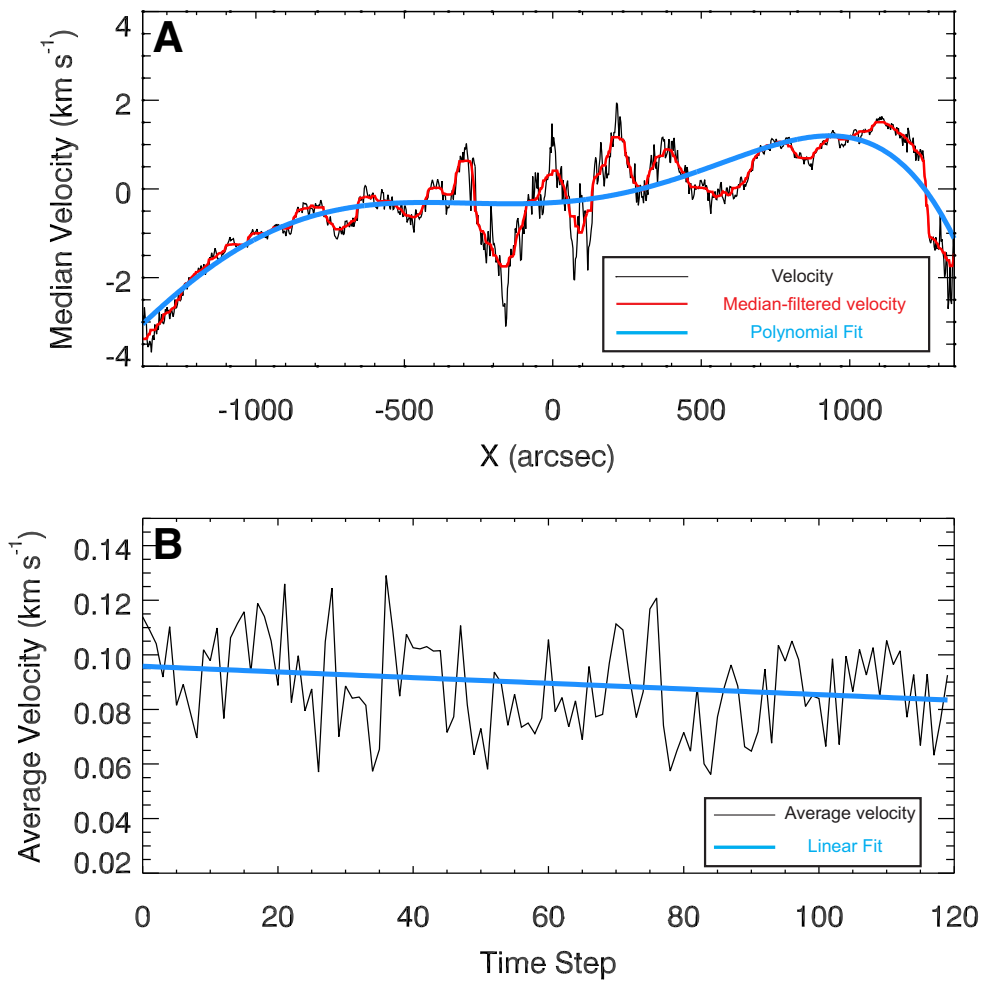

**Fig. S1. Examples of Doppler velocity trend corrections. (A)** The correction for the east-west Doppler velocity trend. This example uses the Doppler velocity image obtained on 19:54:54 UT on 24 February 2022. For every position along X direction (east-west direction), the median value of the Doppler velocity along Y direction (north-south direction) was calculated. The black curve shows the median velocity variation along the east-west direction. The red curve is the median-filtered (in a window of 21 pixels) result of the black curve, with the purpose of removing possible abnormal values. A 5$^{th}$ order polynomial fit was applied to the red curve (shown as the blue curve). This polynomial function describes the east-west variation of the Doppler velocity and will be subtracted from the original Doppler velocity image. **(B)** The correction for the temporal trend of Doppler velocity throughout continuous observation each day. This example shows the correction for the Doppler velocity time series obtained from 19:54:54 UT to 21:01:33 UT on 24 February 2022. For each time step, the average Doppler velocity within the FOV was calculated (black curve). A linear fitting was applied to the time series of the average Doppler velocity to show

potential temporal trend (blue curve), which will be subtracted from the Doppler velocity time series.

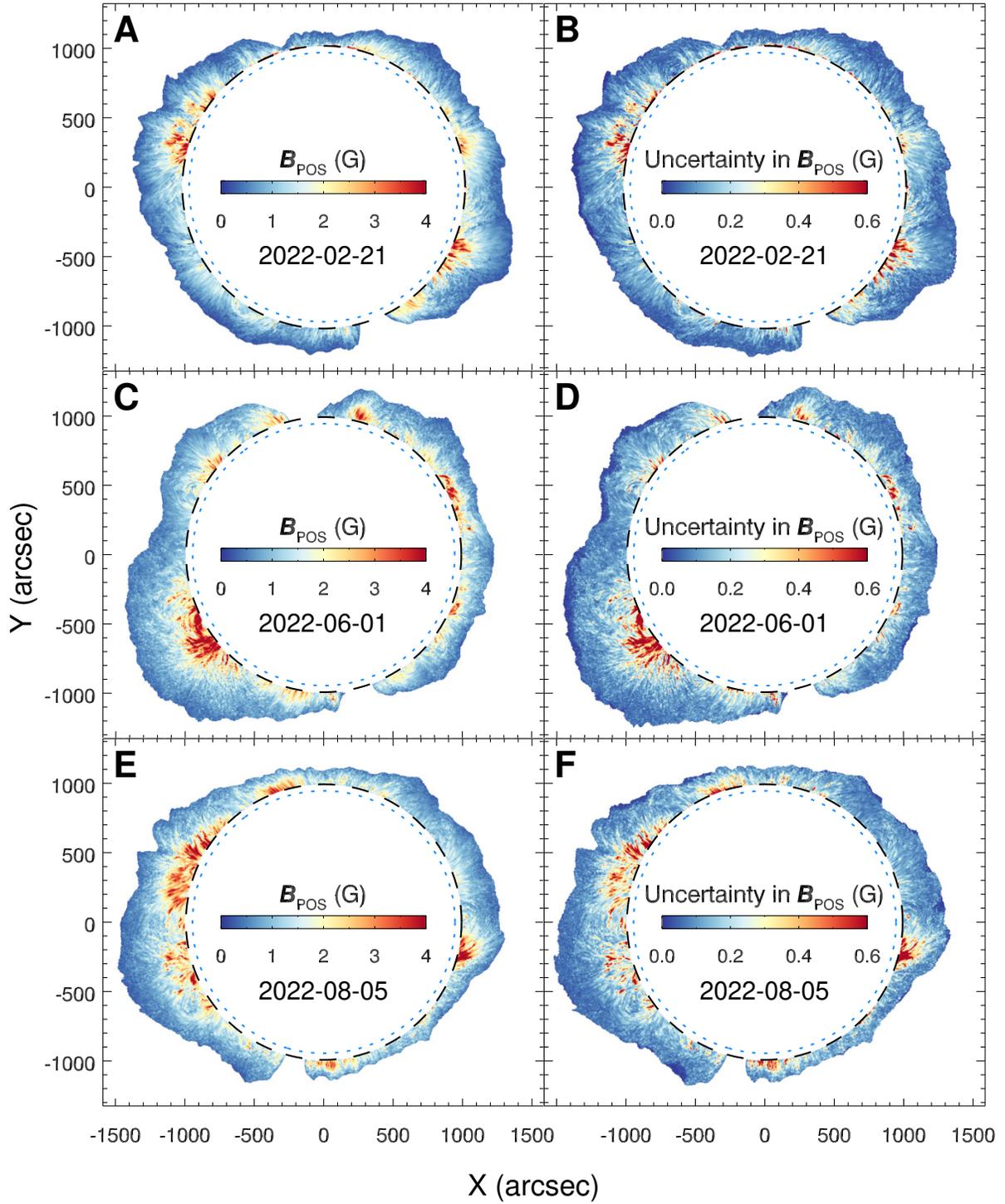

**Fig. S2. Uncertainty on the measured coronal magnetic field strength.** (**A, C, E**) Same as Figure 1B, E, H. (**B, D, F**) The corresponding uncertainties for each of the three measurements, on a different color scale.

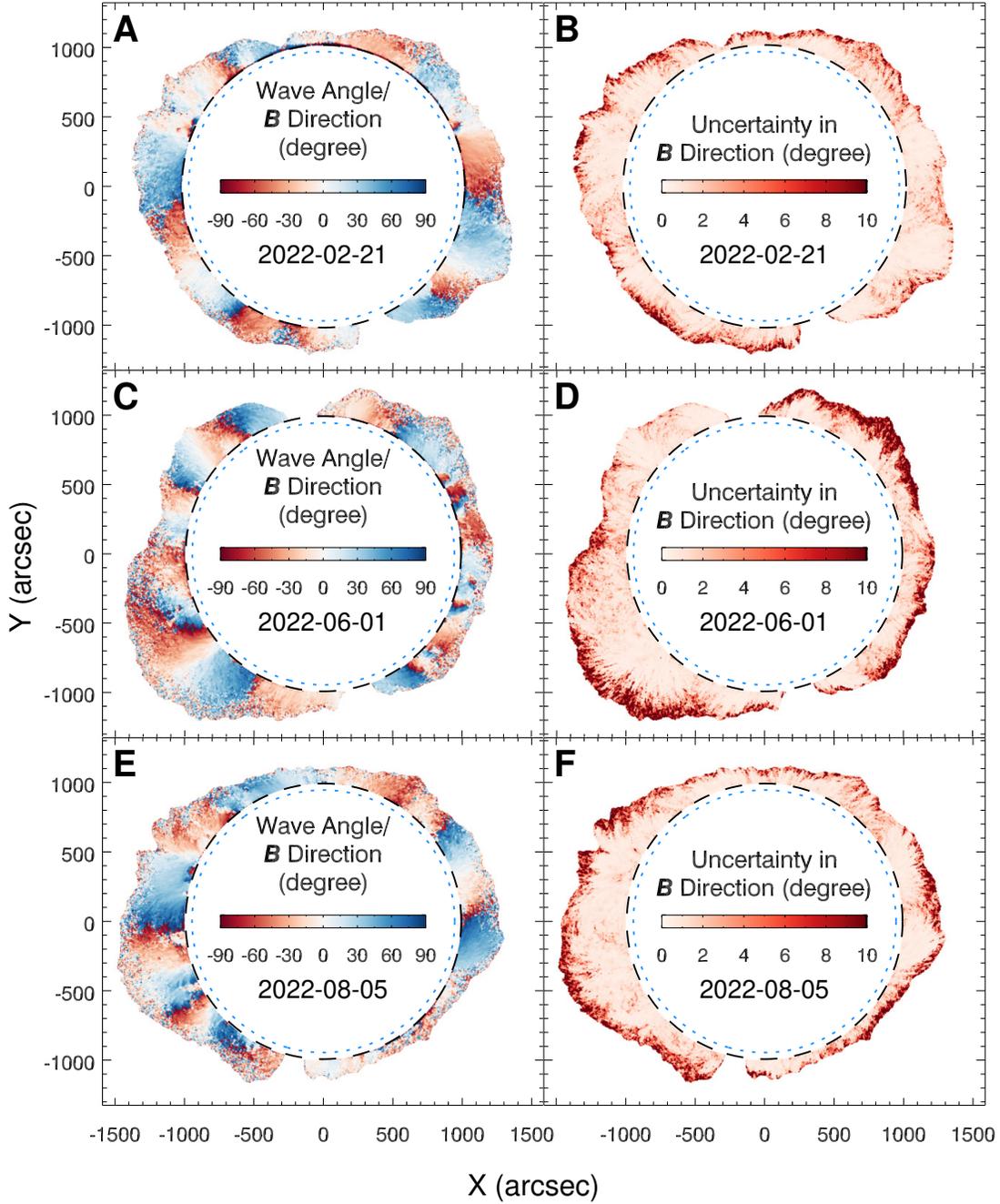

**Fig. S3. Uncertainty on the inferred coronal magnetic field direction.** These uncertainties were determined from the wave propagation angle. (**A, C, E**) Same as Figure 1C, F, I. (**B, D, F**) The corresponding uncertainties for each of the three measurements, on a different color scale.

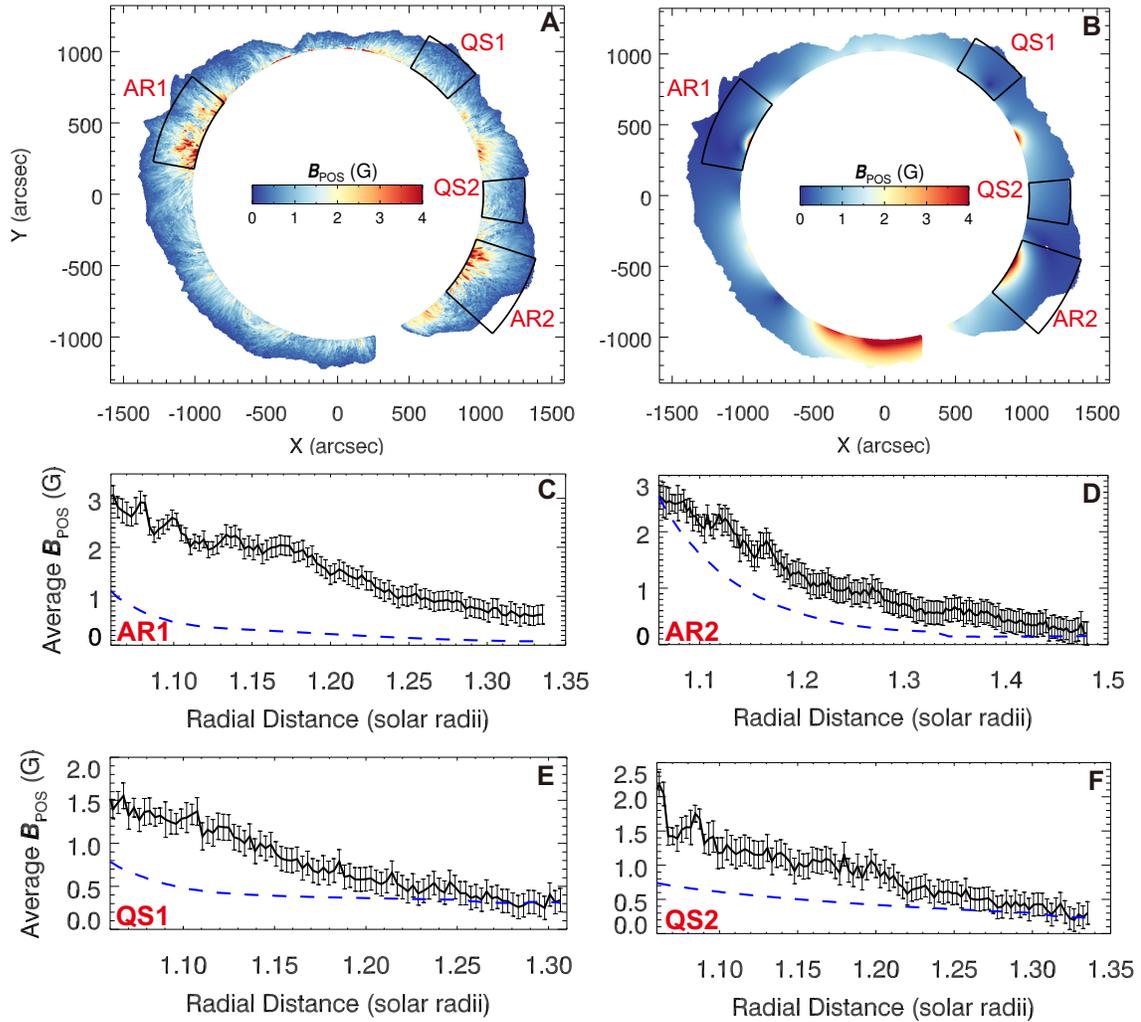

**Fig. S4. Comparison between the UCoMP measurements and PFSS model predictions.** The data are for the UCoMP observations on 21 February 2022 and predictions from a PFSS model using the corresponding HMI synoptic magnetogram as input. **(A)** The same magnetic field strength map as Fig. 1B, but with four annulus sectors overlain marking two active regions (AR1, AR2) and two quiet-Sun regions (QS1, QS2). **(B)** The magnetic field strength on the POS that crosses the solar center predicted by the PFSS model. **(C)-(F)** The average magnetic field strengths as a function of radial distance within the annulus sectors from the UCoMP measurement (black solid line) and PFSS model (blue dashed line). Error bars indicate the standard errors on the mean values in each radius bin.

**Table S1. Log of the UCoMP observations.** We used 114 UCoMP datasets, each consisting of several frames of Fe XIII 1074.7 nm and 1079.8 nm intensity images for plasma density diagnostics, followed by a continuous image sequence used for wave tracking. N indicates the number of frames used for analysis.

| UT Date | Wave tracking | | Fe XIII 1074.7 nm | | Fe XIII 1079.8 nm | |
|---|---|---|---|---|---|---|
| | UT range | N | UT range | N | UT range | N |
| 2022-02-19 | 19:47:04-20:53:42 | 120 | 18:51:15-19:21:39 | 4 | 18:56:34-19:26:57 | 4 |
| 2022-02-20 | 19:51:10-20:57:47 | 120 | 18:55:17-19:25:36 | 4 | 18:00:44-19:31:03 | 4 |
| 2022-02-21 | 19:47:45-20:54:21 | 120 | 18:51:50-19:22:20 | 4 | 18:57:08-19:27:37 | 4 |
| 2022-02-23 | 19:47:30-20:58:43 | 120 | 18:51:45-19:22:04 | 4 | 18:57:02-19:27:22 | 4 |
| 2022-02-24 | 19:54:54-21:01:33 | 120 | 19:26:45-19:29:17 | 2 | 19:32:13-19:34:45 | 2 |
| 2022-02-26 | 19:44:24-20:51:03 | 120 | 18:48:46-19:18:58 | 4 | 18:54:04-19:24:16 | 4 |
| 2022-02-28 | 19:43:53-20:50:35 | 120 | 18:42:31-19:18:27 | 4 | 18:53:34-19:23:45 | 4 |
| 2022-03-01 | 19:29:55-20:35:18 | 115 | 18:29:45-18:59:57 | 4 | 18:35:03-19:05:15 | 4 |
| 2022-03-03 | 19:46:02-20:53:12 | 120 | 18:41:21-19:20:22 | 4 | 18:46:41-19:25:44 | 4 |
| 2022-03-05 | 23:33:42-00:40:19 (+1 day) | 120 | 22:34:40-23:07:28 | 4 | 22:40:05-23:12:54 | 4 |
| 2022-03-06 | 19:36:13-20:42:47 | 120 | 18:40:21-19:10:51 | 4 | 18:45:49-19:16:08 | 4 |
| 2022-03-07 | 19:26:57-20:46:53 | 120 | 18:31:24-19:01:36 | 4 | 18:36:41-19:06:54 | 4 |
| 2022-03-09 | 19:37:33-20:31:17 | 97 | 18:41:41-19:12:11 | 4 | 18:46:59-19:17:29 | 4 |
| 2022-03-12 | 19:13:20-20:19:56 | 120 | 18:17:37-18:47:58 | 4 | 18:22:54-18:53:16 | 4 |
| 2022-03-13 | 19:40:07-20:46:41 | 120 | 18:44:16-19:14:46 | 4 | 18:52:06-19:20:03 | 3 |
| 2022-03-19 | 20:32:58-21:39:36 | 120 | 19:37:05-20:07:26 | 4 | 19:42:22-20:12:53 | 4 |
| 2022-03-20 | 19:08:35-20:15:15 | 120 | 18:11:40-18:43:11 | 4 | 18:16:58-18:48:29 | 4 |
| 2022-03-21 | 02:58:27-04:05:04 | 120 | 02:02:35-02:33:05 | 4 | 02:07:52-02:38:23 | 4 |
| 2022-03-21 | 19:24:18-20:37:07 | 120 | 18:28:33-18:58:56 | 4 | 18:33:51-19:04:14 | 4 |
| 2022-03-24 | 19:06:26-20:13:18 | 120 | 17:56:38-18:55:04 | 4 | 18:02:16-19:00:32 | 4 |
| 2022-03-25 | 20:46:08-21:53:05 | 120 | 19:50:10-20:20:39 | 4 | 19:55:28-20:25:58 | 4 |
| 2022-03-27 | 19:36:40-20:44:13 | 120 | 18:40:47-19:10:58 | 4 | 18:46:05-19:16:25 | 4 |
| 2022-03-31 | 19:16:58-20:23:33 | 120 | 18:21:14-18:51:25 | 4 | 18:26:31-18:56:43 | 4 |
| 2022-04-01 | 19:26:23-20:33:01 | 120 | 18:30:16-19:00:38 | 4 | 18:35:34-19:06:06 | 4 |
| 2022-04-03 | 20:37:41-21:37:02 | 106 | 19:42:10-20:12:20 | 4 | 19:47:27-20:17:38 | 4 |
| 2022-04-10 | 18:54:44-19:25:33 | 56 | 17:58:49-18:29:11 | 4 | 18:04:07-18:34:29 | 4 |
| 2022-04-13 | 19:20:37-20:27:16 | 120 | 18:24:52-18:55:14 | 4 | 18:30:10-19:00:32 | 4 |
| 2022-04-24 | 20:37:28-21:43:28 | 120 | 19:28:49-20:12:50 | 4 | 19:34:04-20:18:05 | 4 |
| 2022-04-26 | 00:21:52-01:27:50 | 120 | 23:26:45 (-1 day) - 23:56:41 (-1 day) | 4 | 23:32:00 (-1 day) - 00:01:56 | 4 |
| 2022-04-27 | 20:34:30-21:40:33 | 120 | 19:35:07-19:37:39 | 2 | 20:12:03-20:14:34 | 2 |
| 2022-05-13 | 19:26:04-20:32:06 | 120 | 18:30:41-19:00:51 | 4 | 18:35:57-19:06:07 | 4 |
| 2022-05-14 | 19:53:27-20:54:20 | 108 | 18:50:09-19:20:08 | 4 | 18:55:24-19:25:23 | 4 |
| 2022-05-21 | 19:56:10-21:02:13 | 120 | 19:00:50-19:30:59 | 4 | 19:06:06-19:36:14 | 4 |
| 2022-05-22 | 19:41:39-20:47:42 | 120 | 18:45:01-19:16:15 | 4 | 18:51:21-19:21:40 | 4 |
| 2022-05-23 | 18:56:36-20:02:38 | 120 | 18:00:09-18:30:16 | 4 | 18:05:24-18:36:31 | 4 |
| 2022-06-01 | 18:34:06-19:40:09 | 120 | 17:38:45-18:08:54 | 4 | 17:44:00-18:14:10 | 4 |
| 2022-06-02 | 19:07:57-20:13:58 | 120 | 18:12:47-18:42:45 | 4 | 18:18:02-18:48:01 | 4 |
| 2022-06-04 | 18:38:21-19:44:22 | 120 | 17:43:03-18:13:00 | 4 | 17:48:18-18:18:26 | 4 |
| 2022-06-06 | 19:04:29-20:10:30 | 120 | 18:09:17-18:39:16 | 4 | 18:14:32-18:44:31 | 4 |
| 2022-06-07 | 18:36:50-19:42:52 | 120 | 17:41:29-18:11:28 | 4 | 17:46:44-18:16:44 | 4 |
| 2022-06-08 | 18:36:40-19:42:42 | 120 | 17:41:20-18:11:29 | 4 | 17:46:35-18:16:44 | 4 |
| 2022-06-09 | 18:42:12-19:48:14 | 120 | 17:47:01-18:17:00 | 4 | 17:52:16-18:22:16 | 4 |
| 2022-06-10 | 21:01:40-22:07:42 | 120 | 20:33:57-20:36:28 | 2 | 20:39:12-20:41:44 | 2 |
| 2022-06-11 | 18:44:54-19:50:55 | 120 | 17:49:35-18:19:43 | 4 | 17:55:00-18:24:58 | 4 |
| 2022-06-13 | 00:58:34-02:04:36 | 120 | 00:03:05-00:33:14 | 4 | 00:10:51-00:38:29 | 3 |
| 2022-06-13 | 19:38:19-20:44:23 | 120 | 18:43:00-19:13:07 | 4 | 18:48:15-19:18:23 | 4 |
| 2022-06-14 | 18:41:56-19:47:58 | 120 | 17:46:20-18:16:30 | 4 | 17:51:35-18:21:55 | 4 |
| 2022-06-16 | 18:52:50-19:58:54 | 120 | 17:56:07-18:26:11 | 4 | 18:01:25-18:31:27 | 4 |
| 2022-06-19 | 18:51:46-19:57:50 | 120 | 17:56:00-18:26:10 | 4 | 18:01:16-18:31:45 | 4 |
| 2022-06-20 | 18:42:40-19:48:43 | 120 | 17:47:27-18:17:26 | 4 | 17:52:42-18:22:41 | 4 |
| 2022-06-21 | 18:46:14-19:52:15 | 120 | 17:51:00-18:21:01 | 4 | 17:56:16-18:26:16 | 4 |
| 2022-06-24 | 18:59:58-20:06:00 | 120 | 18:04:45-18:34:45 | 4 | 19:10:01-18:40:01 | 4 |
| 2022-06-25 | 18:42:45-19:39:54 | 104 | 17:47:22-18:17:32 | 4 | 17:52:38-18:22:48 | 4 |
| 2022-06-27 | 22:13:12-23:18:40 | 119 | 21:45:28-21:48:00 | 2 | 21:50:44-21:53:15 | 2 |
| 2022-07-03 | 19:36:02-20:16:34 | 74 | 18:39:20-10:09:20 | 4 | 18:44:35-19:16:06 | 3 |
| 2022-07-04 | 20:40:35-21:46:37 | 120 | 20:12:52-20:15:53 | 2 | 20:18:07-20:20:39 | 2 |
| 2022-07-05 | 21:07:33-22:13:36 | 120 | 20:08:06-20:42:22 | 4 | 20:13:21-20:47:37 | 4 |
| 2022-07-07 | 19:26:07-20:18:50 | 95 | 18:30:56-19:00:56 | 4 | 18:36:12-19:06:12 | 4 |
| 2022-07-09 | 18:49:19-19:55:21 | 120 | 17:53:42-18:23:42 | 4 | 17:58:59-18:29:07 | 4 |
| 2022-07-10 | 19:27:07-20:33:09 | 120 | 18:31:46-19:01:56 | 4 | 18:37:01-19:07:11 | 4 |
| 2022-07-12 | 18:52:14-19:58:17 | 120 | 17:56:45-18:26:53 | 4 | 18:02:10-18:32:08 | 4 |
| 2022-07-14 | 19:08:52-20:14:53 | 120 | 18:13:11-18:43:39 | 4 | 18:18:26-18:48:56 | 4 |
| 2022-07-17 | 20:00:16-21:06:18 | 120 | 19:04:36-19:34:55 | 4 | 19:09:53-19:40:20 | 4 |
| 2022-07-18 | 18:55:31-19:28:49 | 61 | 17:49:24-18:19:31 | 4 | 17:54:39-18:24:57 | 4 |
| 2022-07-20 | 19:18:31-20:24:34 | 120 | 18:23:09-18:53:19 | 4 | 18:28:25-18:58:34 | 4 |
| 2022-07-21 | 19:25:28-20:31:31 | 120 | 18:20:52-18:57:06 | 4 | 18:26:17-19:02:22 | 4 |
| 2022-07-22 | 19:16:48-20:23:28 | 120 | 18:11:53-18:48:13 | 4 | 18:17:11-18:53:31 | 3 |
| 2022-07-24 | 19:08:21-20:14:59 | 120 | 18:03:26-18:39:47 | 4 | 18:08:44-18:45:05 | 4 |
| 2022-07-25 | 19:35:08-20:39:00 | 115 | 18:25:54-19:06:30 | 4 | 18:31:13-19:11:48 | 4 |
| 2022-07-26 | 19:31:42-20:38:19 | 120 | 18:26:27-19:02:58 | 4 | 18:31:45-19:08:16 | 4 |
| 2022-07-27 | 19:21:03-20:27:41 | 120 | 18:09:36-18:50:20 | 4 | 18:14:54-18:55:38 | 4 |
| 2022-07-28 | 19:39:23-20:12:59 | 61 | 18:27:57-19:08:42 | 4 | 18:33:15-19:14:00 | 4 |
| 2022-07-30 | 21:42:45-22:18:02 | 64 | 20:31:06-21:11:51 | 4 | 20:36:24-21:17:18 | 4 |
| 2022-07-31 | 19:30:17-20:33:35 | 114 | 18:18:50-18:59:35 | 4 | 18:24:08-19:04:53 | 4 |
| 2022-08-01 | 19:50:37-20:41:01 | 91 | 18:32:00-19:19:55 | 4 | 18:37:18-19:25:13 | 4 |
| 2022-08-02 | 19:36:30-20:30:49 | 98 | 18:24:53-19:05:38 | 4 | 18:30:11-19:10:56 | 4 |
| 2022-08-05 | 19:31:16-20:37:54 | 120 | 18:20:00-19:00:36 | 4 | 18:25:18-19:05:53 | 4 |
| 2022-08-06 | 19:29:04-20:21:10 | 94 | 18:17:39-18:58:24 | 4 | 18:22:47-19:03:42 | 4 |
| 2022-08-14 | 19:43:55-20:50:34 | 120 | 18:30:20-19:11:54 | 4 | 18:35:48-19:17:22 | 4 |
| 2022-08-19 | 19:49:30-20:22:10 | 59 | 18:36:23-19:16:58 | 4 | 18:41:41-19:22:15 | 4 |
| 2022-08-23 | 19:49:31-20:56:12 | 120 | 18:38:14-19:18:49 | 4 | 18:43:32-19:24:07 | 4 |
| 2022-08-24 | 19:32:36-20:09:38 | 60 | 18:21:18-19:01:55 | 4 | 18:26:36-19:07:13 | 4 |
| 2022-08-25 | 20:02:01-21:08:43 | 120 | 18:49:49-19:30:25 | 4 | 18:55:07-19:35:43 | 4 |
| 2022-08-27 | 19:32:31-20:17:53 | 82 | 18:20:54-19:01:40 | 4 | 18:26:12-19:06:58 | 4 |
| 2022-08-28 | 19:37:35-20:44:15 | 120 | 18:26:00-19:06:44 | 4 | 18:31:18-19:12:02 | 4 |
| 2022-08-29 | 20:06:26-20:56:56 | 89 | 18:32:15-19:14:49 | 4 | 18:37:33-19:20:07 | 4 |
| 2022-08-30 | 20:07:09-21:08:11 | 110 | 18:51:40-19:36:28 | 4 | 18:56:58-19:41:46 | 4 |
| 2022-08-31 | 19:38:08-20:44:47 | 120 | 18:26:51-19:07:27 | 4 | 18:32:09-19:12:45 | 4 |
| 2022-09-01 | 20:09:58-21:16:36 | 120 | 18:58:24-19:39:08 | 4 | 19:03:52-19:44:26 | 4 |

| 2022-09-04 | 19:35:21-20:42:02 | 120 | 18:23:56-18:26:28 | 2 | 18:29:14-18:31:46 | 2 |
| 2022-09-05 | 20:32:29-21:17:18 | 81 | 19:37:27-19:39:59 | 2 | 19:42:44-19:45:17 | 2 |
| 2022-09-06 | 20:01:11-20:30:52 | 54 | 18:38:37-19:19:10 | 4 | 18:43:55-19:24:28 | 4 |
| 2022-09-08 | 19:48:46-20:55:25 | 120 | 19:14:18-19:16:50 | 2 | 19:19:35-19:22:08 | 2 |
| 2022-09-09 | 20:04:32-21:11:19 | 109 | 19:30:47-19:33:19 | 2 | 19:36:04-19:38:37 | 2 |
| 2022-09-10 | 21:21:49-22:03:48 | 76 | 19:08:11-20:13:07 | 4 | 19:13:28-20:18:24 | 4 |
| 2022-09-12 | 21:33:44-22:28:02 | 98 | 19:33:51-19:36:23 | 2 | 19:39:09-19:41:41 | 2 |
| 2022-09-14 | 19:37:06-20:35:55 | 95 | 19:03:21-19:05:54 | 2 | 19:08:39-19:11:11 | 2 |
| 2022-09-17 | 20:24:23-21:13:14 | 88 | 18:55:06-19:35:49 | 4 | 19:00:23-19:41:06 | 4 |
| 2022-09-18 | 19:42:49-20:46:06 | 114 | 18:28:04-19:08:48 | 4 | 18:33:21-19:14:06 | 4 |
| 2022-09-19 | 20:03:00-21:00:42 | 104 | 19:16:32-19:19:04 | 2 | 19:22:39-19:25:12 | 2 |
| 2022-09-20 | 19:51:12-20:50:33 | 107 | 18:38:38-19:16:38 | 3 | 18:41:23-19:21:56 | 4 |
| 2022-09-22 | 19:58:01-20:53:26 | 98 | 17:59:18-18:39:51 | 4 | 18:04:35-18:42:37 | 3 |
| 2022-09-27 | 20:11:15-21:05:35 | 98 | 18:54:57-19:35:30 | 2 | 19:02:47-19:40:47 | 3 |
| 2022-09-29 | 20:43:16-21:12:23 | 53 | 19:27:29-19:30:01 | 2 | 19:32:46-19:35:19 | 2 |
| 2022-10-12 | 20:39:09-21:28:59 | 90 | 19:23:31-20:04:12 | 4 | 19:28:48-20:09:30 | 4 |
| 2022-10-15 | 20:01:16-21:07:57 | 120 | 18:49:41-19:30:36 | 4 | 18:55:09-19:35:53 | 4 |
| 2022-10-16 | 19:52:00-20:31:47 | 60 | 19:17:30-19:20:03 | 2 | 19:22:48-19:25:20 | 2 |
| 2022-10-17 | 19:59:03-20:34:21 | 64 | 18:47:37-19:28:22 | 4 | 18:52:55-19:33:40 | 3 |
| 2022-10-18 | 20:23:11-21:24:48 | 111 | 19:03:25-19:44:00 | 4 | 19:08:43-19:49:18 | 4 |
| 2022-10-20 | 20:01:22-20:40:34 | 71 | 18:52:30-19:30:42 | 3 | 18:55:16-19:36:00 | 4 |
| 2022-10-23 | 20:10:27-21:01:25 | 92 | 19:21:33-19:24:05 | 2 | 19:26:51-19:29:23 | 2 |
| 2022-10-25 | 20:14:53-21:21:32 | 120 | 19:06:10-19:44:12 | 3 | 19:08:55-19:49:30 | 4 |
| 2022-10-26 | 19:48:33-20:28:19 | 72 | 18:37:08-19:17:53 | 4 | 18:42:35-19:23:11 | 4 |
| 2022-10-29 | 20:34:05-21:14:26 | 72 | 18:38:55-18:41:28 | 2 | 18:44:13-18:46:46 | 2 |

**Caption for Movie S1 (.mp4 file).**

The 3.5-mHz filtered Doppler velocity images observed on 21 February 2022.

**Caption for Movie S2 (.mp4 file).**

All 114 global maps of the coronal magnetic field measured from the UCoMP observations. Plotting symbols are the same as Fig. 1B-C. Note that we obtained two coronal magnetograms on 21 March 2022, and also on 13 June 2022 (from the two datasets observed on each day, see Table S1). For these two days we use "(1)" and "(2)" to identify the two magnetograms.

**Caption for Movie S3 (.mp4 file).**

The spherical distributions of magnetic field strength in different atmospheric layers. An animated version of Figure 3, showing all meridians over five Carrington rotations. Plotting symbols and color scales are the same as in Fig. 3. (**A**) The photospheric magnetic field (radial component) measured by HMI. (**B**) The coronal magnetic field averaged between 1.10 and 1.15 solar radii derived from the UCoMP observations. (**C**) Same as panel B, except averaged between 1.20 and 1.25 solar radii.